\documentclass[preprints,article,accept,moreauthors,pdftex]{Definitions/mdpi}

\newcommand{\HI}{{\ion{H}{1}}}

\newcommand{\kms}{$\,$km$\,$s$^{-1}$}
\newcommand{\WHz}{$\,$W$\,$Hz$^{-1}$}
\newcommand{\ergs}{$\,$erg$\,$s$^{-1}$}

\newcommand{\mJybeam}{mJy beam$^{-1}$}

\newcommand{\msun}{{$M_\odot$}}
\newcommand{\msunyr}{{$M_\odot$ yr$^{-1}$}}
\newcommand{\dotm}{{$\dot{\rm M}$}}

\newcommand{\cc}{\mbox{cm$^{-3}$}}

\newcommand{\coOne}{{CO(1-0)}}
\newcommand{\coTwo}{{CO(2-1)}}

\newcommand{\alwmath}[1]{\ifmmode#1\else$#1$\fi}

\def\HI{H{\,\small I}}

\newcommand{\ltsima} {$\; \buildrel < \over \sim \;$}
\newcommand{\gtsima} {$\; \buildrel > \over \sim \;$}
\newcommand{\lta} {\lower.5ex\hbox{\ltsima}}
\newcommand{\gta} {\lower.5ex\hbox{\gtsima}}

\firstpage{1} 
\pubvolume{1}
\issuenum{1}
\articlenumber{0}
\pubyear{2022}
\copyrightyear{2022}
\externaleditor{{Academic Editor: Stefi Baum and Christopher P. O'Dea } 
}
\datereceived{12 December 2022} 
\daterevised{26 January 2023} 
\dateaccepted{30 January 2023} 
\datepublished{} 
\hreflink{https://doi.org/} 

\Title{Young Radio Sources Expanding in Gas-Rich ISM: Using Cold Molecular Gas to Trace Their Impact}

\TitleCitation{Young Radio Sources Expanding in Gas-Rich ISM: Using Cold Molecular Gas to Trace Their Impact}

\Author{Raffaella Morganti 
 $^{1,2,}$*\orcidA{}, Suma Murthy $^{3}$, Pierre Guillard $^{4,5}$, Tom Oosterloo $^{1,2}$\orcidB{} and Santiago Garcia-Burillo $^{6}$}

\AuthorNames{Raffaella Morganti, Suma Murthy, Pierre Guillard, Tom Oosterloo and Santiago Garcia-Burillo}

\AuthorCitation{Morganti, R.; Murthy, S.; Guillard, P.; Oosterloo, T.; Garcia-Burillo, S.}

\address{%
$^{1}$ \quad ASTRON, 
 The Netherlands Institute for Radio Astronomy, Oude Hoogeveensedijk 4, 7991 PD Dwingeloo, 
 The Netherlands; oosterloo@astron.nl 
\\
$^{2}$ \quad Kapteyn Astronomical Institute, University of Groningen, Landleven 12, \mbox{9747 AD Groningen, The Netherlands}  \\
$^{3}$ \quad Joint Institute for VLBI ERIC, Oude Hoogeveensedijk 4, \mbox{7991 PD Dwingeloo, The~Netherlands;} murthy@jive.eu \\
$^4$ \quad Sorbonne Universit\'{e}, CNRS, UMR 7095, Institut d'Astrophysique de Paris, 98bis bd Arago, \mbox{75014 Paris, France}; guillard@iap.fr \\
$^5$ \quad Institut Universitaire de France, Minist{\`e}re de l'Enseignement Sup{\'e}rieur et de la Recherche, 1 rue Descartes, CEDEX 05, \mbox{75231 Paris, France} \\
$^6$ \quad Observatorio Astron\'omico Nacional (OAN-IGN)-Observatorio de Madrid, Alfonso XII, 3, \mbox{28014 Madrid, Spain}; s.gburillo@oan.es
}
\corres{Correspondence: morganti@astron.nl}

\abstract{We present an overview of the results obtained from the study of the resolved distribution of molecular gas around  eight young ($\lta$$10^6$ yr), peaked-spectrum radio galaxies. Tracing the distribution and kinematics of the gas around these radio sources allows us to trace the interplay between the jets and the surrounding medium. For three of these sources, we present new  \coOne\ observations, obtained with the Northern Extended Millimeter Array (NOEMA) with arcsecond resolution. In two of these targets, we also detected CN lines, both in emission and absorption. Combining the new observations with already published data, we discuss the main results obtained. Although we found that a large fraction of the cold molecular gas was distributed in disc-like rotating structures, in the vast majority of the sources, high turbulence and deviations from purely quiescent gas (including outflows) were observed in the region co-spatial with the radio continuum emission. This suggests the presence of an interaction  between radio plasma and cold molecular gas. In particular, we found that  newly born and young radio jets, even those with low power i.e.,\ $P_{\rm jet}<10^{45}$ \ergs),  
are able to drive massive outflows of cold, molecular gas. The outflows are, however, limited to the sub-kpc regions and likely short lived. 
On larger scales (a few kpc), we observed cases where the molecular gas appears to avoid the radio lobes and, instead, wraps around them. The results suggest the presence of an evolutionary sequence, which is consistent with previous simulations, where the type of impact of the radio plasma changes as the jet expands, going from a direct jet-cloud interaction able to drive gas outflows on sub-kpc scales to a more gentle pushing aside of the gas, increasing its turbulence and likely limiting its cooling on kpc scales. This effect can be mediated by the cocoon of shocked gas inflated by the jet--cloud interactions. Building larger samples of young and evolved radio sources for observation at a similar depth and spatial resolution to test this scenario is now needed and may be possible thanks to more data becoming available in the growing public archives. }

\begin{document}

\section{Introduction}

Tracing the properties of the gas in the centre of galaxies has provided important insights into the interplay between the energy emitted by an active galactic nucleus (AGN) and the surrounding medium. Two highly intertwined processes, fuelling and feedback, play a main role (e.g.,\ \cite{Gaspari20,Storchi19}) and are thought to be crucial in the evolution of the host galaxy.
Among the various types of AGNs, radio (jetted)
 AGNs are known to have jets and lobes ranging in size from pc to hundreds of kpc. Spanning such large scales implies that this type of AGN can potentially  have an impact on the gas in the interstellar medium (ISM), circum-galactic medium (CGM), and intergalactic medium (IGM). They are, therefore, particularly interesting for the study of AGN feedback if~the radio plasma efficiently couples with this gas. Although the effect of jets and lobes on larger, intergalactic scales is well established, thanks to  studies of X-ray cavities (\cite{Fabian12,McNamara12}), it is now becoming clear that radio jets can also impact the ISM inside the host galaxy. As suggested by numerical simulations (e.g.,~\cite{Sutherland07}), this can occur in different ways, from an initial phase with a strong interaction (possibly driving fast and massive outflows) to~a more rapid, momentum-dominated expansion phase when the jet breaks free from the denser central regions and reaches galactic sizes and~beyond. 

\begin{figure}[H]
\begin{adjustwidth}{-\extralength}{0cm}
\centering
\includegraphics[width=15.3cm]{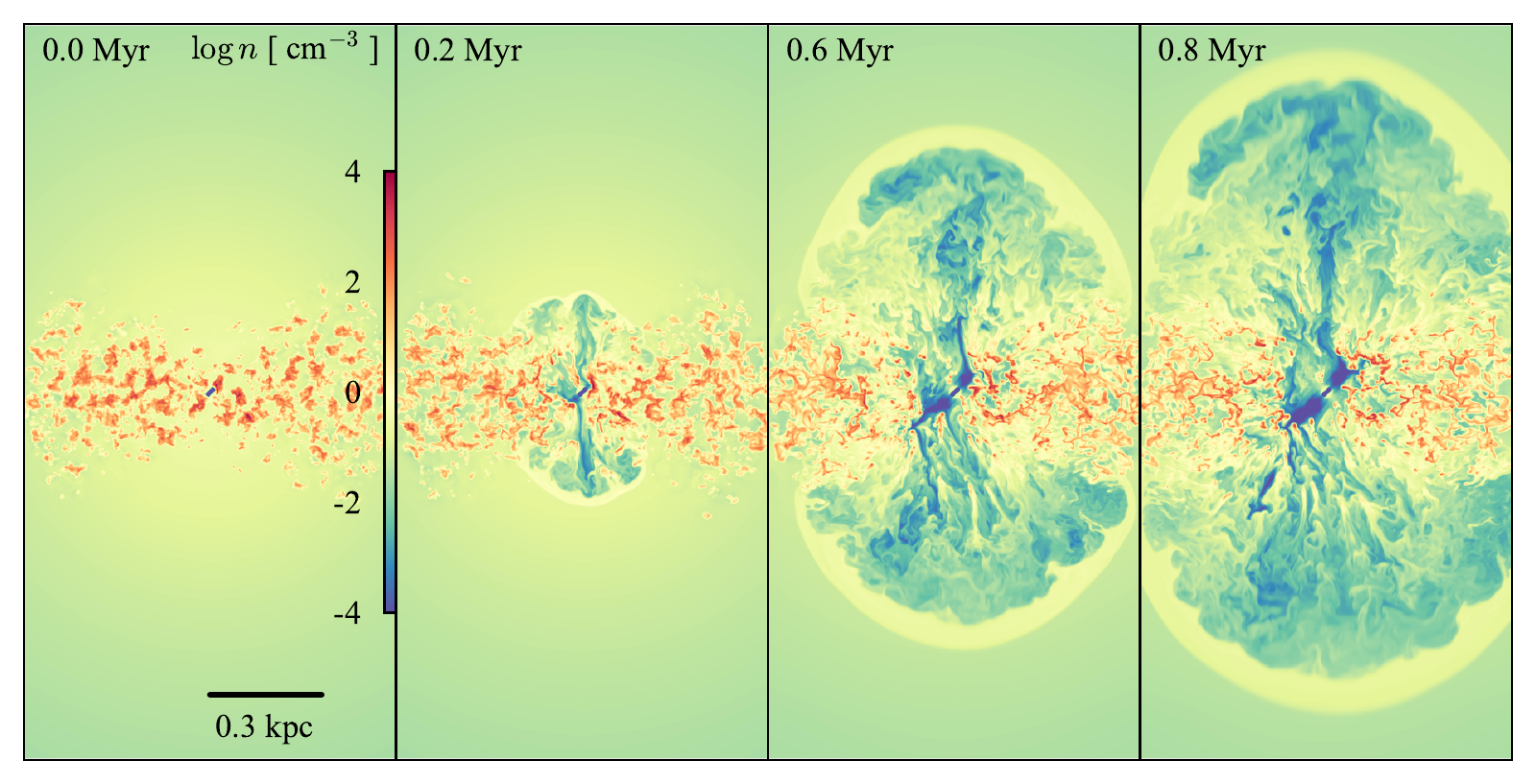}
 \end{adjustwidth}
\caption{\label{fig:simulations} Density 
 slices (logarithmic) from the simulations of~\cite{Mukherjee18}, showing the evolution of the jet--disc system in~the case of a jet slightly offset compared to the disc. The~figure is taken from~\cite{Murthy22} and describes the case of the radio galaxy B2~0258+35.  The~jet plasma is shown in blue and the dense ISM clouds in orange-red. The ablated gas and shocked ambient medium are in yellow. The~strongest interactions occur within the disc where the main jet hits the clouds head-on. These regions show enhanced velocity dispersion and bulk velocities of a few hundred \kms\  and are locations of sharp jet deflection and splitting. The~outer disc is dispersed but~remains largely intact (see~\cite{Murthy22,Mukherjee18}).}

\end{figure}

The actual impact will, however, depend on several parameters: the jet power, orientation of the jet with respect to the gaseous structure, and evolutionary stage of the jet. This complexity is well described by the results from recent numerical simulations (e.g.,\ \cite{Sutherland07,Wagner12,Mukherjee18,Perucho21,Talbot22,Cielo18}).
In particular, the~simulations developed by~\cite{Sutherland07,Wagner12} and~more recently expanded by~\cite{Mukherjee18} illustrate the process of a jet entering a clumpy medium and initially directly interacting with  the surrounding gas clouds. They also show how the impact of the jet is amplified by the creation of a cocoon of a shocked medium resulting from the jet--cloud interactions and~expanding orthogonal to the jet (see Figure\  \ref{fig:simulations} for an example).  This latter phase starts in the galactic ISM and is also visible as an X-ray emission around the radio source (e.g.,\ \cite{Croston07,Croston09,Mingo11,Mingo12}), bearing similarities with the X-ray cavities observed in the~IGM.

Some numerical simulations are also starting to provide direct comparisons to observations~\cite{Mukherjee18-IC5063,Murthy22}. An~example illustrating how an interaction proceeds is shown in Figure~\ref{fig:simulations}, where simulations recently proposed to describe the results from the cold molecular gas for the young radio galaxy B2~0258+35 are shown~\cite{Murthy22} (see also Section~\ref{sec:publishedResults} for details). 

\textls[-11]{Thus, both the simulations and several observational results suggest that the initial phase ($<$$10^6$ yr;~\cite{Fanti90,Snellen04,ODea21}) of the evolution of the radio jet, with~the jet expanding through the inner (few) kpc of the host galaxy,  can be particularly important. Tracing the gas around the young radio jets is, therefore, key to quantifying the effects of this kind of interaction, which were described above. This has been carried out extensively for the  warm ionised gas (\cite{Holt09,Shih13,Santoro20}).} However, the~cold phase of the gas --- and, in particular, the cold molecular gas that is used in this paper ---is highly relevant for its direct connection to star formation. Furthermore, it can be observed with the high spatial resolution necessary for tracing this~interaction.

Here, we provide an overview of the results obtained so far by an ongoing project aimed at tracing the properties of the cold gas in young radio galaxies embedded in a rich gaseous medium. 
We focus on the properties of the cold molecular gas component traced with sensitive, high spatial resolution observations carried out with the Northern Extended Millimeter Array (NOEMA) and the Atacama Large Millimeter Array (ALMA). The high spatial resolution obtained for most of the targets has been used to resolve the distribution and kinematics of the gas across the radio emission and to better describe the effects (if any) of the  jet--ISM~interaction. 

This paper is structured as follows. In Section~\ref{sec:observations}, we describe the selection of the target group of objects of our observations and our study; in Section~\ref{sec:observations}, we present the as-yet unpublished  \coOne\ results for the three objects; in Section\ \ref{sec:publishedResults}, we summarise the published results for the rest of the sample; in Section~\ref{sec:generalResults}, we combine our findings and highlight the main points, and in Section\ \ref{sec:discussionFuture}, we finish with some conclusions about the implications of the results obtained so far and remarks about future~work.

Throughout this paper we adopt a cosmology that assumes a flat Universe and the following parameters: $H_{\circ} = 70$ \kms\ Mpc$^{-1}$, $\Omega_\Lambda = 0.7$, $\Omega_{\rm M} = 0.3$.

\vspace{-6pt}

\section{Cold Molecular Gas in Young Radio Sources: Overview of Our~Observations }

The eight sources observed in CO are listed in Table~\ref{tab:PropertiesTargets}, together with a selection  of their relevant properties. 
The list  includes known young 
 radio galaxies (i.e.,\ small and peaked spectrum sources) residing in a gas-rich medium. In~particular, the~selected objects have all been  detected in \HI\ absorption, ensuring the presence of cold gas~\cite{Maccagni18,Holt08,Gereb15,Morganti05a,Murthy22,Struve12}. 

The targets are all small radio sources of linear size  $\lesssim  4$ kpc, with~a peaked radio spectrum and a turnover frequency ranging from about 100 MHz to a few GHz. These are properties typical of radio galaxies recently born (i.e.,\ $\lesssim$10$^6$ yr old; see~\cite{ODea21} for a review). In~all cases, VLBI observations of their radio continuum structures are available in the~literature.

The original goal of the project was to build a sample covering a broad range of source ages and jet powers (using the radio power as a proxy to a first approximation) to be able to investigate the trends proposed by the numerical simulations. Although~the final list of observed objects is still limited, the~observed objects cover a relatively large range in these parameters. In more detail, the~ages cover a range from 91 yr (obtained from  hotspot advance measurements~\cite{Giroletti09}) in the case of PKS~1718--63, which is  considered one of the youngest known radio sources, to~ 3.5 Myr for 3C~305~\cite{Hardcastle12}, the~oldest in the sample. Additionally, the radio power (proxy for jet power) of the sources in our sample covers a large range, as seen in Table~\ref{tab:PropertiesTargets}, with~PKS~1718--63, IC~5063, and B2~0258+35 at the lower end of the distribution and PKS~0023--26 and PKS~1549--79 the most powerful sources in the sample. These powerful sources also host powerful radiative AGNs~\cite{Holt09,Santoro20}. The~presence of both radiative and mechanical energy from these AGNs makes them good candidates for tracing powerful feedback effects. For more details, we also refer to the papers listed in Table~\ref{tab:PropertiesTargets}. 
   
Of the eight sources, in~five of them, the cold molecular gas  has been extensively discussed (see references in Table~\ref{tab:PropertiesTargets} for details and Section\ \ref{sec:publishedResults} for a short overview of recent results). For the three remaining sources, in  Section~\ref{sec:observations}, we present the results from new NOEMA \coOne\ observations. For one object (4C~31.04), a paper is currently being prepared that includes, together with CO, new \HI\ VLBI observations (Murthy~et~al.\ in prep). As a result, here, we present only the highlights of the CO results relevant to the discussion in this~paper. 

\begin{table}[H]

\caption{\label{tab:PropertiesTargets}Properties of the~targets.}

\begin{adjustwidth}{-\extralength}{0cm}
\setlength{\cellWidtha}{\fulllength/8-2\tabcolsep+0in}
\setlength{\cellWidthb}{\fulllength/8-2\tabcolsep-0.1in}
\setlength{\cellWidthc}{\fulllength/8-2\tabcolsep-0.1in}
\setlength{\cellWidthd}{\fulllength/8-2\tabcolsep+0in}
\setlength{\cellWidthe}{\fulllength/8-2\tabcolsep-0.1in}
\setlength{\cellWidthf}{\fulllength/8-2\tabcolsep-0.1in}
\setlength{\cellWidthg}{\fulllength/8-2\tabcolsep+0.5in}
\setlength{\cellWidthh}{\fulllength/8-2\tabcolsep-0.1in}
\scalebox{1}[1]{\begin{tabularx}{\fulllength}{>{\centering\arraybackslash}m{\cellWidtha}>{\centering\arraybackslash}m{\cellWidthb}>{\centering\arraybackslash}m{\cellWidthc}>{\centering\arraybackslash}m{\cellWidthd}>{\centering\arraybackslash}m{\cellWidthe}>{\centering\arraybackslash}m{\cellWidthf}>{\centering\arraybackslash}m{\cellWidthg}>{\centering\arraybackslash}m{\cellWidthh}}

\toprule
\multicolumn{1}{l}{\textbf{Target}} & \textbf{Telescope} & \textbf{Transition} & \textbf{Beam}    &  \textbf{log}\boldmath{$L_{\rm 1.4~GHz}$} &  \textbf{Size Radio} &\textbf{ Structure  Molecular}  &  \textbf{References} \\
       &           &            & \textbf{Arcsec} {\bf (kpc)}  & \textbf{\WHz}  &   \textbf{kpc}        &  \textbf{Gas Distribution} &  \\
\midrule 
\multicolumn{1}{l}{PKS~1718--64}
 & ALMA & CO(2-1) & $0.28 \times 0.19$ & 24.23 &  0.002 &  disc & \cite{Maccagni18}  
 \\  
\multicolumn{1}{l}{(NGC6328)} & &  & ($0.08 \times 0.06$)  & &    &   infall clouds &    \\
\midrule
\multicolumn{1}{l}{IC5063} & 	ALMA& CO(1-0) & $0.54\times 0.45$ & 23.73 & 0.93  & 	disc  & 	\cite{Morganti15,Oosterloo17} \\
 & & CO(2-1) & ($0.12 \times 0.10$)& & &  outflow & \\
 & & CO(3-2) &  & & &   &  \\
\midrule
\multicolumn{1}{l}{B2~0258+35} &	NOEMA &	CO(1-0) &	$1.9\times 1.5$ & 23.32 &	 1.2 &	 	(disc) &	\cite{Murthy22} \\
  & &   & ($0.64 \times 0.50$) & & &  outflow & \\
 \midrule
\multicolumn{1}{l}{PKS~1549-79} &	ALMA&	CO(1-0) &  	$0.17\times 0.13$ &	26.39 &	 0.4 &	disc &  \cite{Oosterloo19} \\
  & &  & ($0.45\times 0.34$) &  & &  outflow & \\ 
  & & CO(3-2) & $0.31\times 0.16$ &  & &   & \\ 
    & &  & ($0.82\times 0.42$) &  & & &   \\ 
   \midrule
\multicolumn{1}{l}{PKS0023-26} &	ALMA &	CO(2-1) &	$0.45\times 0.34$ &  
	27.42 &	4.3 &	 diffuse  &	 \cite{Morganti21-0023} \\
 &  & & ($2.10\times 1.59$) & &  &  expand shell&  \\
\midrule
\multicolumn{1}{l}{3C305} & 	NOEMA	& CO(1-0) &	$1.52 \times 0.89$&	25.03 &	4.1 & disc	& this paper  \\
    & &  & ($1.25\times 0.73$) &  & & &    \\
    \midrule
\multicolumn{1}{l}{4C52.37}&	NOEMA&	CO(1-0) & $2.02 \times 1.74$ 	&	25.28 &	0.35 &	disc &	this paper \\
    & &  & ($3.92\times 3.38$) &  & & &   \\
    \midrule
\multicolumn{1}{l}{4C31.04}	& NOEMA &	CO(1-0) &	$2.09 \times 1.34$ &	25.29	& 0.14	& disc	&  this paper \\
  & &  & ($2.42 \times 1.55$) & &   & expand shell &  \\
  
\bottomrule 
\end{tabularx} }
 \end{adjustwidth} 
 \noindent\footnotesize{{\bf The columns are:} (1) source name;  (2) telescopes used; (3) beam of the CO observations (in parenthesis, the scale in kpc); (4) the radio luminosity at 1.4~GHz; (5) the size of the radio sources (in the case of B2~0258+35, only the central, young source is given, see~\cite{Brienza18} for details); (6) the structure of the distribution of the cold molecular gas, as described in the paper given in the~references. 
 }
\end{table}

\vspace{-6pt}

As can be seen in Table~\ref{tab:PropertiesTargets}, all  sources have been observed in either CO(1-0) or CO(2-1). For only two cases, data on more than one CO transition are available. The availability of more than one transition has been key to identifying the signatures of the interplay between the jet and the ISM,  for example, by identifying gas with a high excitation temperature. The observations have spatial resolutions ranging from 0.2 to 2.0 arcsec, allowing us to spatially resolve the distribution of CO across the radio continuum emission in most cases.  

In Section~\ref{sec:observations}, we start by describing the new results from the NOEMA observations of 3C~305, 4C~52.37, and 4C~31.04,  as well as providing a short description of the general properties of these three objects. In~Section\ \ref{sec:publishedResults}, we present a summary of  the CO observations of the remaining objects in the sample, including the results obtained at other wavebands relevant to the final interpretation of the CO~data.  

\section{New NOEMA \coOne\ Observations of 3C~305, 4C~52.37, and 4C~31.04}
\label{sec:observations}

Three targets, 3C 305, 4C 52.37, and 4C~31.04, were observed with the NOEMA (in the intermediate C configuration) under project W18CR004 for the first two sources and S21BP001 for the latter source. For~3C~305 and  4C~52.37, the~observations were carried out over three days using nine to ten antennas. The~setup consisted of two sidebands, each 7.4 GHz wide and~subdivided into 3859 channels, giving a spectral resolution of 2 MHz  ($\sim$ 5--6 \kms). In the case of 4C~52.37, the~lower and the upper sidebands were centred at 90.4 GHz and 105.6 GHz, respectively. In~the case of 3C~305, the~central frequencies were 96.7 GHz and 112.4 GHz, respectively. We used 3C 345 and 3C 454.4 as flux calibrators and MWC349, J1637+574, J1629+495, J1458+718, J1418+546, and J1604+572 as phase~calibrators.  

The observations of 4C~31.04 were carried out with a similar setup to the two sidebands centred at 94.7 GHz and 110.2 GHz, respectively (Murthy~et~al.\ in prep).


The data reduction followed the standard steps. The initial flagging of  bad visibilities and the calibration were performed using the  Grenoble Image and Line Data Analysis Software (GILDAS). Next, the~two polarisations were averaged and the calibrated visibilities exported to uvfits format for further reduction in CASA. The~data were  initially self-calibrated on the continuum emission  via a few cycles of imaging and phase-only calibration, followed by a round of amplitude-and-phase calibration. The~line-free channels were then used to fit a first-order polynomial to each calibrated visibility spectrum and subtracted from the \textit{uv} data. Cubes were produced with a robust parameter of --1 and natural weighting in order to characterise the distribution and kinematics of the molecular gas, ensuring that sure low-surface brightness emission could be detected. The parameters of the final cubes are summarised in Table~\ref{tab:Observations1}.

\begin{table}[H]
\caption{Observation~details.}
\begin{adjustwidth}{-\extralength}{0cm}
\setlength{\cellWidtha}{\fulllength/9-2\tabcolsep-0.1in}
\setlength{\cellWidthb}{\fulllength/9-2\tabcolsep-0.2in}
\setlength{\cellWidthc}{\fulllength/9-2\tabcolsep-0.1in}
\setlength{\cellWidthd}{\fulllength/9-2\tabcolsep+0in}
\setlength{\cellWidthe}{\fulllength/9-2\tabcolsep+0.2in}
\setlength{\cellWidthf}{\fulllength/9-2\tabcolsep+0.1in}
\setlength{\cellWidthg}{\fulllength/9-2\tabcolsep-0.1in}
\setlength{\cellWidthh}{\fulllength/9-2\tabcolsep+0.3in}
\setlength{\cellWidthi}{\fulllength/9-2\tabcolsep-0.1in}
\scalebox{1}[1]{\begin{tabularx}{\fulllength}{>{\centering\arraybackslash}m{\cellWidtha}>{\centering\arraybackslash}m{\cellWidthb}>{\centering\arraybackslash}m{\cellWidthc}>{\centering\arraybackslash}m{\cellWidthd}>{\centering\arraybackslash}m{\cellWidthe}>{\centering\arraybackslash}m{\cellWidthf}>{\centering\arraybackslash}m{\cellWidthg}>{\centering\arraybackslash}m{\cellWidthh}>{\centering\arraybackslash}m{\cellWidthi}}

\toprule
\textbf{Source} & \boldmath{$\Delta t$}  & \textbf{beam}\boldmath{$_{\rm rob}$}  & \textbf{PA}\boldmath{$_{\rm rob}$} & \textbf{RMS}\boldmath{$_{\rm rob}$} & \textbf{beam}\boldmath{$_{\rm nat}$}  & \textbf{PA}\boldmath{$_{\rm nat}$} & \textbf{RMS}\boldmath{$_{\rm nat}$} & \boldmath{$\Delta v $}\\
 & \textbf{(h) }& \boldmath{$(`` \times$}\textbf{'')} & \boldmath{$(^{\circ})$} & \textbf{(mJy beam}\boldmath{$^{-1}$}\textbf{)} & \boldmath{$(`` \times$}\textbf{'')} & \boldmath{$(^{\circ})$} & \textbf{(mJy beam}\boldmath{$^{-1}$}\textbf{)} & \textbf{\kms}\\
 \textbf{(1)} & \textbf{(2)} & \textbf{(3) }&\textbf{ (4)} & \textbf{(5)} &\textbf{ (6)} &\textbf{ (7) }& \textbf{(8)} &\textbf{(9)} 
 \\ 
\midrule
3C~305    & 5.6  &  \mbox{1.52 $\times$ 0.89} & 11.52 & 0.6 & 1.99 $\times$ 1.18 & 16.52 & 0.5 & 21.6\\
4C~52.37  & 3.4  &  \mbox{2.02 $\times$ 1.74} & 66.44 & 0.4 & 2.30 $\times$ 2.01 & $-75.7$ & 0.35 & 23.0\\
4C~31.04
 & 9.8 & \mbox{2.09 $\times$ 1.34} & 16.28 & 0.5 & 2.43 $\times$ 1.56 & 17.50 & 0.3 & 27.5 \\
\bottomrule
\end{tabularx}   }
\end{adjustwidth}
\noindent\footnotesize{{\bf The columns are:} (1) source name;  (2) on-source time; (3), (4), (5) beam, position angle, and the RMS of the cube made with robust $=-1$ weighting; (6), (7), (8) beam, position angle, and the RMS of the cube made with natural weighting; (9) velocity resolution.}

\label{tab:Observations1}
\end{table}

\vspace{-6pt}

We de-redshifted the \textit{uv} data to the systemic velocity of the target using the redshifts listed in Table~\ref{tab:Observations} and obtained the final spectral cubes by imaging these datasets after averaging four channels. The~velocity resolutions of the cubes are listed in Table~\ref{tab:Observations}.

The cubes were visually inspected for the presence of \coOne. However, to~ensure that the analysis was systematic, they were also run through the Source Finding Application (SoFiA;~\cite{Serra15})\endnote{https://github.com/SoFiA-Admin/SoFiA.}. This software performs spatial and velocity smoothing in order to optimise the search for~emission.

In order to derive the molecular mass, we first estimated the CO(1-0) luminosity $L^\prime_{\rm CO(1-0)}$ \citep{Solomon05,Bolatto13}:
\begin{equation}
L^\prime_{\rm CO(1-0)} =2453 \  D^2_{\rm L}\ (1+z)^{-1}\ S_{\rm CO(1-0)}\Delta\nu
\end{equation}
where $z$ is the redshift, $D_{\rm L}$  the luminosity distance in Mpc,  and~$S_{\rm CO(1-0)}\Delta v$  the integrated  flux in Jy \kms\ of the CO(1-0) line. 
The molecular mass was estimated using $M_{\rm H_2}=\alpha_{\rm CO}  L^\prime_{\rm CO(1-0)}$, 
where $\alpha_{\rm CO}$ is the CO-to-H$_2$ conversion  factor in units of \msun (K \kms\ pc$^2$)$^{-1}$.
The derived masses of the molecular gas are given in Table~\ref{tab:Observations} using $\alpha_{\rm CO} = 3.6 \ M_\odot/({\rm K\ km\ s^{-1}\ pc^2})$ \cite{Bolatto13}.
This choice of $\alpha_{\rm CO}$ is based on the fact that, as~discussed below, a~large fraction of gas is regularly rotating. However, this is not the case for all the regions (in particular in 3C~305 and 4C~31.04, see below) and, therefore, a~different (lower) conversion factor could be required. Thus, the~values of the H$_2$ mass listed in Table~\ref{tab:Observations} should be considered the upper limits.

The  3 mm continuum images were obtained from the line-free channels and were made with the same weightings as those used for the line~cubes. 

\begin{table}[H]
\caption{Molecular gas~masses.}
\newcolumntype{C}{>{\centering\arraybackslash}X}
\begin{tabularx}{\textwidth}{CCCC}\toprule
\textbf{Source} & \boldmath{$z$} & \textbf{S}\boldmath{$_{CO} \times \Delta v $} & \textbf{M}\boldmath{$_{\rm H_2}$} \\
 &  & (Jy \kms)  & ($\times 10^{9}$ \msun)\\
 (1) & (2) & (3) & (4)\\ 
\midrule
3C~305    & 0.0417 & 7.39  & 2.3\\
4C~52.37  & 0.106  & 5.81  & 12\\
4C~31.04 & 0.0602  & 3.64 & 2.4 \\
\bottomrule
\end{tabularx}

\noindent\footnotesize{The columns are (1) source name, (2) redshift, (3) molecular gas mass estimated, assuming $\alpha_{\rm CO} = 3.6$\ M$_\odot/({\rm K\ km\ s^{-1}\ pc^2})$. See text for details.}

\label{tab:Observations}
\end{table}

\vspace{-6pt}

The width of the observing band allows us to explore the presence of the CN(1-0) lines. Interestingly, these were detected (in emission and absorption) in two of the targets (4C~52.37 and 4C~31.04). We did not detect CN(1-0) in 3C~305, possibly due to the limited sensitivity of these observations. The~discussion of the CN lines is presented in Section~\ref{sec:CN}.

In the following sections, we describe the results obtained from the \coOne\ observations of the three~targets.



\subsection{3C~305}
\label{sec:3c305}

\begin{figure}[H]
\includegraphics[width=0.7\textwidth]{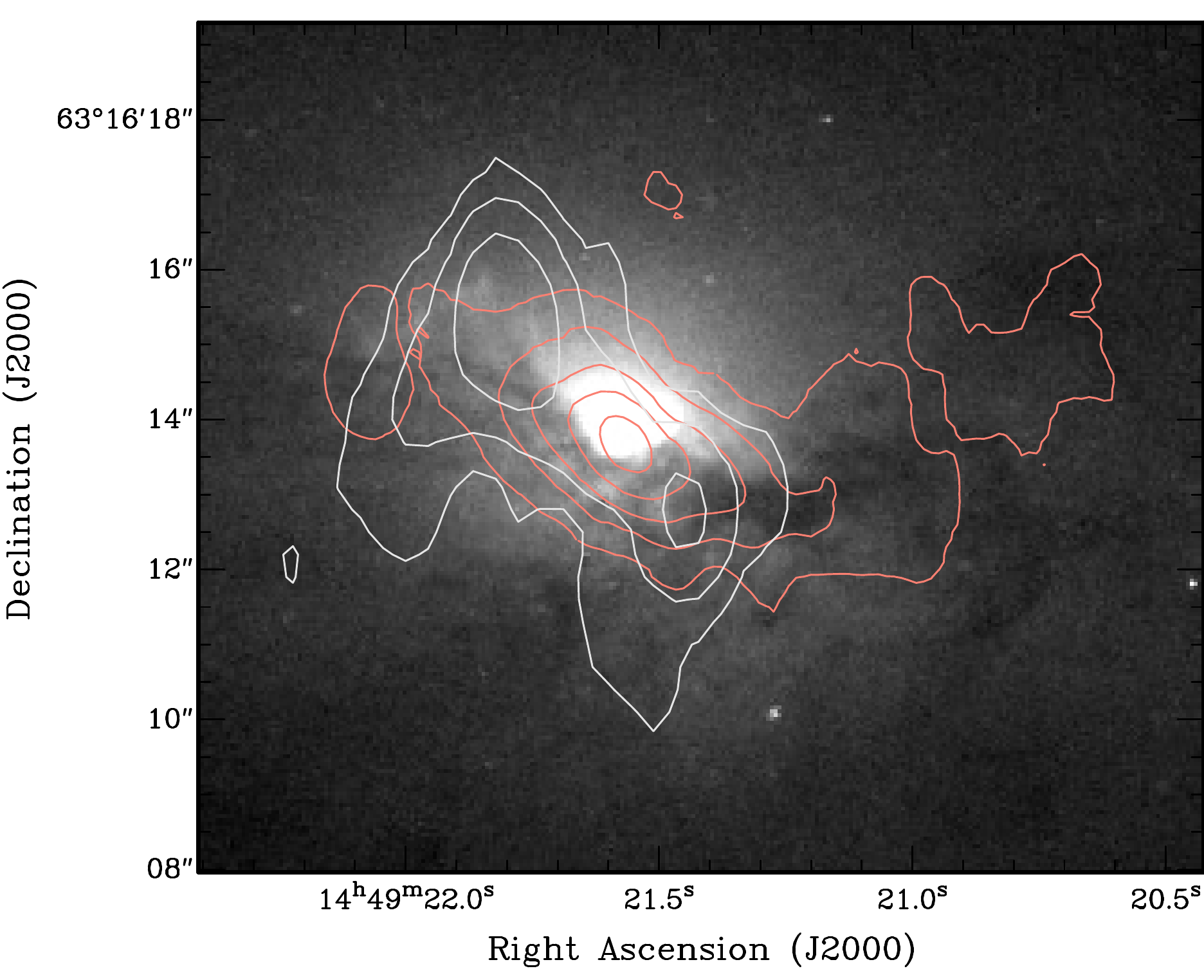}
\caption{\label{fig:3C305overlay} HST WFPC2 image of 3C~305 in the F555 filter, showing the inner dust lane with~superimposed \coOne\ total intensity contours (orange) and 3 mm continuum (white). The~molecular gas clearly follows the dust lane. Contour levels for the continuum are 0.20, 0.84, and~3.53 \mJybeam and for the \coOne\ total intensity are 0.14, 0.44, 0.74, 1.04, and~1.34 Jy beam$^{-1}$\kms.}
\end{figure}

This radio galaxy and the multi-phase gas in which it is embedded have been the topic of several studies~\cite{Jackson03,Morganti05,Hardcastle12,Guillard12}. The~radio continuum morphology is peculiar, with~two bright jets and hotspots, as well as lobes, which extend mostly perpendicular to the radio axis (see~\cite{Hardcastle12} and the references therein, and~Figure~\ref{fig:3C305overlay}). This structure suggests the presence of an interaction between the radio plasma and the ISM. Indeed, such an interaction has been traced, particularly on the E side, by~the kinematics and properties of the ionised and  \HI\ gas, which show broad lines and outflows (\cite{Jackson03,Morganti05}). The~physical conditions and kinematics of the warm molecular gas have been investigated by~\cite{Guillard12} using the {\em  Spitzer Infrared Spectrograph (IRS)}. This revealed multiple warm H$_2$ lines  with FWHM above 500 \kms, indicating that the warm H$_2$ gas is also very turbulent in this source. In~line with this and~based on the X-ray morphology and spectral properties derived with Chandra,~\cite{Hardcastle12} argued that the extended X-ray emission was likely produced by shock-heating from interaction with the radio~jet.

\begin{figure}[H]
\begin{adjustwidth}{-\extralength}{0cm}
\centering
\includegraphics[height=0.3\textwidth]{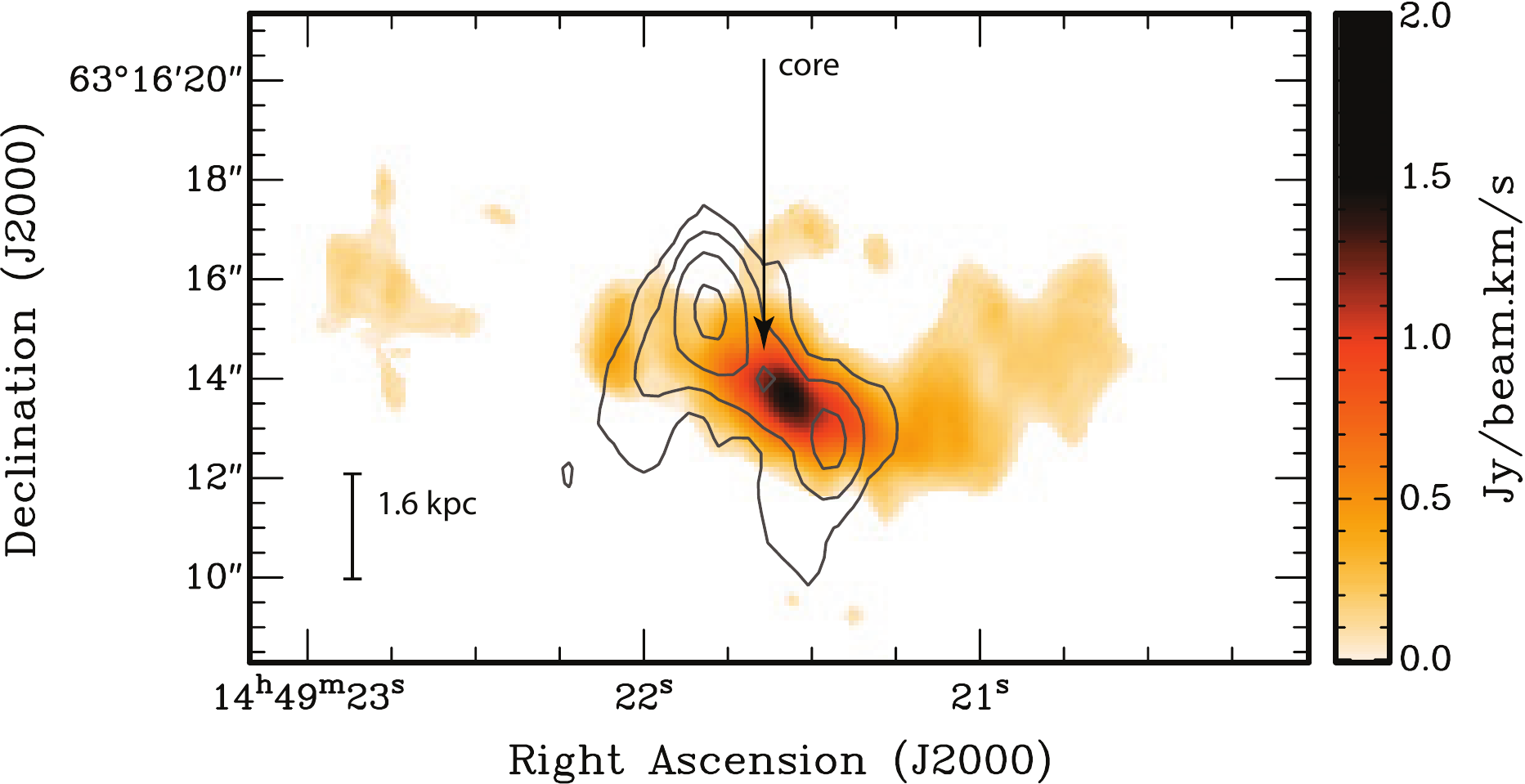} 
\hskip0.5cm
\includegraphics[height=0.3\textwidth]{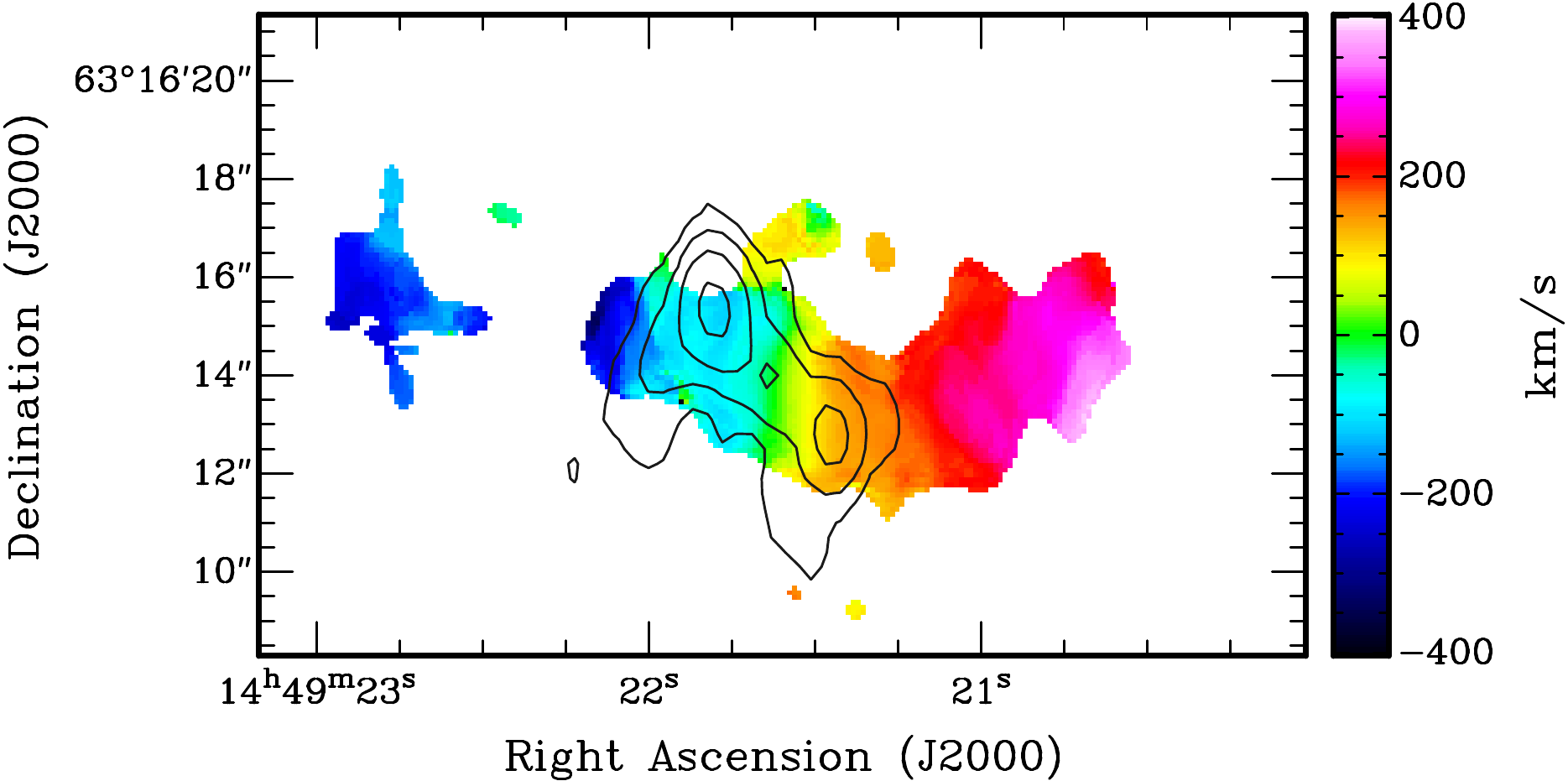} \\
\includegraphics[height=0.5\textwidth]{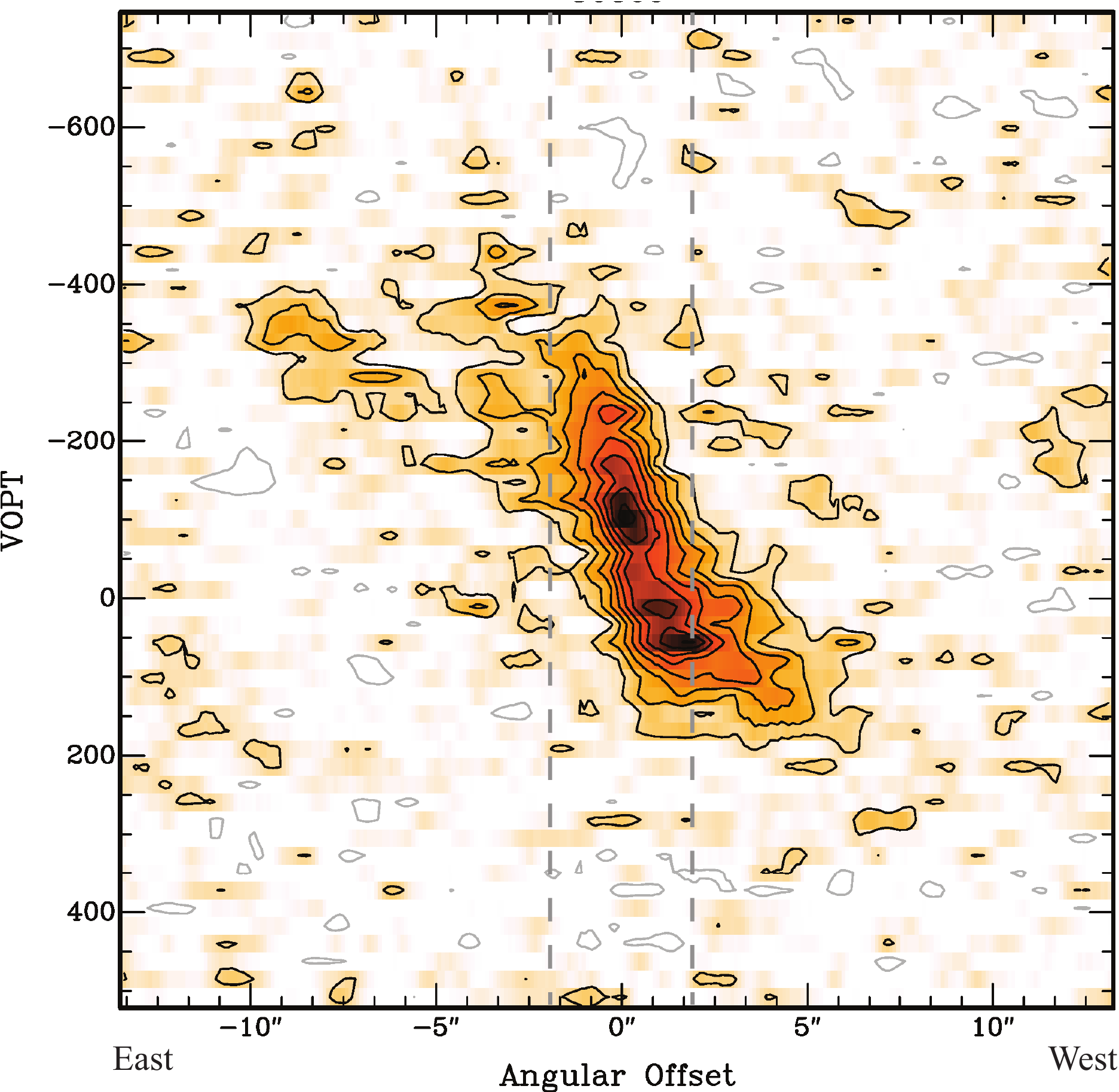}
\includegraphics[height=0.5\textwidth]{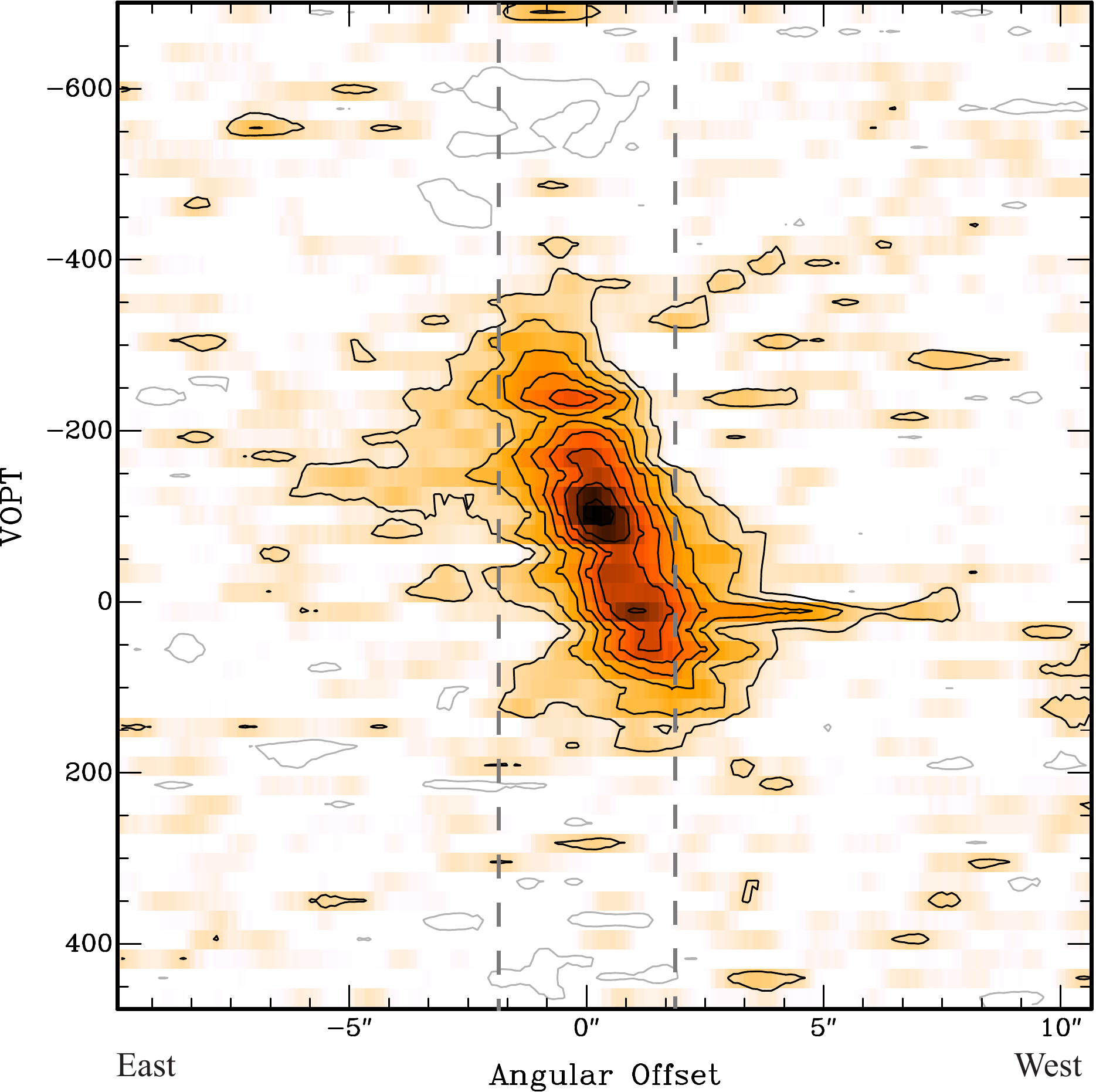}
\end{adjustwidth}
\caption{Top: 3C~305 moment-0 (left) and moment-1 (right) maps obtained using a mask to include only emission above the 4-$\sigma$ level. This was made from the cube with robust weighting (see Table~\ref{tab:Observations}). The~contours show the radio continuum map at a similar spatial resolution. Contour levels in both top panels are 0.2, 0.8, 2.8, and~10.0 \mJybeam. The~location of the radio core is indicated.
Bottom: position--velocity plot along the major axis (left) of the distribution of the molecular gas and (right) along the radio axis. The vertical dashed lines indicate the extent of the radio source. Contour levels in both position--velocity diagrams are 1, 2, 3, 4, 5, 6, 7, and~  8 \mJybeam.}
\label{fig:3C305}
\end{figure}

Cold molecular gas was earlier detected in 3C~305 using single-dish observations  with the IRAM-30m telescope~\cite{Evans05,Ocana10}. 
Our new NOEMA observations have the angular resolution to trace the distribution of the gas. The~molecular gas is distributed in a rotating structure roughly aligned with the radio axis and follows the warped dust lane, as~shown in Figure~\ref{fig:3C305overlay} (orange contours). 
The total intensity (moment-0) image and the velocity field (moment-1) image overlaid with the contours from the 3 mm continuum emission are shown in the top panels of Figure~\ref{fig:3C305}. As can be seen in the moment-0 image (where the location of the radio core is also indicated), the~total extent of the distribution of the molecular gas is about 17 arcsec, corresponding to 14 kpc. A~brighter, inner region extending to $\sim$4 arcsec (about 3 kpc) can be observed. This is mostly co-spatial with the inner radio continuum emission, whereas the brightness decreases at the location of the hotspots (see below). The~molecular gas extends on the W side to about 8 arcsec (6.6 kpc) following a warped structure. On~the E side, most of the molecular gas extends to 4 arcsec, whereas more but~fainter emission is detected up to 9 arcsec (7.4 kpc).  The~velocity field shows the overall rotation, which dominates the CO structure, and even the cloud east of 3C 305 follows these kinematics. However, at~radii larger than $\sim$2 arcsec (1.6 kpc),~in particular, on the E side, the~gas distribution  and kinematics  of the molecular gas appear to be less regular.  This is clearer in the  position--velocity  plots obtained along the major axis of the molecular gas along the radio axis (Figure~\ref{fig:3C305} bottom). The~ dashed vertical lines indicate the locations of the hotspots in the radio continuum, showing the broad emission ($>$150 \kms) seen, in particular, on the E side.

We derived a total molecular gas mass of $2.3 \times 10^9$ \msun\ using a Galactic CO-H$_2$ conversion factor of $\alpha_{\rm CO} = 3.6$ M$_\odot/({\rm K\ km\ s^{-1}\ pc^2})$. This is consistent with that derived by~\cite{Ocana10}, who found $2.1 \times 10^9$ \msun\ of molecular gas. However, since we also found that in some regions, the gas showed  high-velocity dispersion (pointing to optically thin gas, see below), a lower conversion factor may be more appropriate for those regions. In~this case, the estimated molecular gas mass represents the upper~limit.

Interestingly, the~location of the broad emission is also where \HI\ and ionised gas outflows were detected and could represent the molecular counterpart of these outflows. The~broad CO emission is too faint for a proper characterisation of the kinematics of this gas. Nevertheless, if~we assume that this gas represents the counterpart of the \HI\ outflow, some useful constraints on the properties of a possible molecular outflow can be derived. We estimated the upper limit of the mass of this potential molecular outflow by considering a 3-$\sigma$ signal over 400 \kms\  (i.e.,\ a  width comparable to the outflow observed in \HI). Using a conversion factor of  $\alpha_{\rm CO} = 0.34$ $M_\odot/({\rm K\ km\ s^{-1}\ pc^2})$, typical of  the optically thin regime found in outflows, we derived an upper limit  of $M_{\rm H_2}^{\rm out}$$\sim$$ 1.5\times 10^7$ \msun. The~corresponding mass outflow rate was \dotm $\sim$4 \msunyr. Considering the large uncertainties due to the assumptions made, this is consistent with the \HI\ mass outflow rate estimated in~\cite{Morganti05}, suggesting that the molecular outflowing component, if~present, is not~large. 
 
A direct comparison between the \HI\ and CO outflows is complicated by the fact that the \HI\ outflow has been observed in the absorption and thus can only be traced against the regions where the radio continuum is detected.  The~molecular gas seen in the emission suggests that the situation may be more complex. For~example, we note that the faint emission of CO gas with broad profiles is mostly located  east of the peak of the continuum (indicated by the dashed line in Figure~\ref{fig:3C305}). Additionally, it is interesting that the total intensity image (Figure\  \ref{fig:3C305} top left) suggests that the molecular gas (especially on the E side)  avoids the brighter part of the radio lobe. Thus, in~the case of 3C~305, we may be seeing the radio lobe in the process of clearing the molecular gas by pushing it aside. This could be similar to what was seen in PKS~0023--26. There, the~molecular gas with broad, albeit not extremely broad, profiles was seen wrapping around the lobe (see~\cite{Morganti21-0023} for details). We revisit this in the discussion in Section\ \ref{sec:generalResults}.

\subsection{4C~52.37}
\label{sec:4c52.37}

4C~52.37 is a radio galaxy that is included in the CORALZ  sample  of  compact  and  young radio sources selected by~\cite{Snellen04}. In~the  VLBI sub-arcsecond radio morphology, it has been classified as a compact symmetric object~\cite{Vries09}.  

\begin{figure}[H]
\centering
\begin{adjustwidth}{-\extralength}{0cm}
\includegraphics[height=0.3\textwidth]{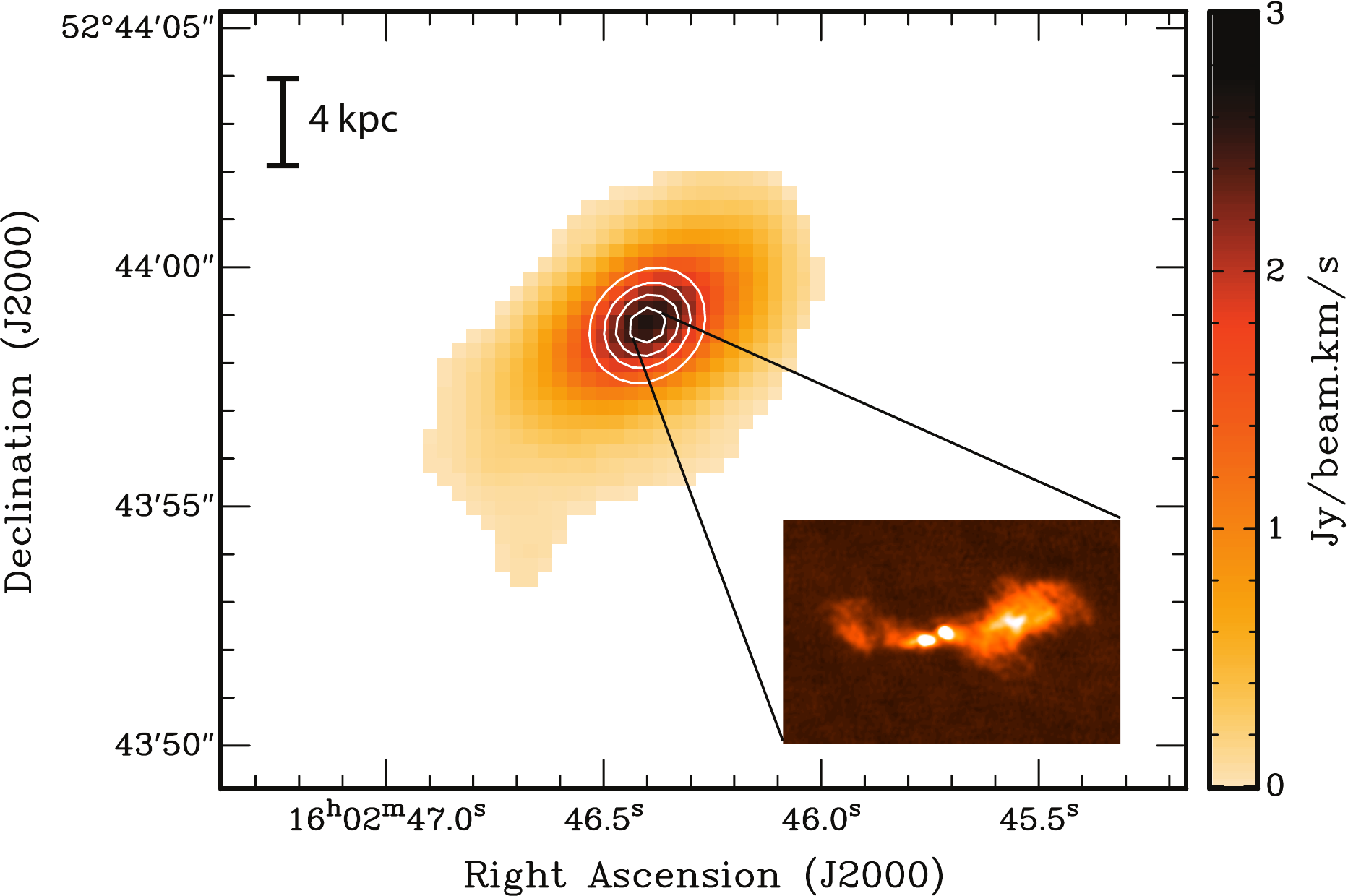}
\includegraphics[height=0.3\textwidth]{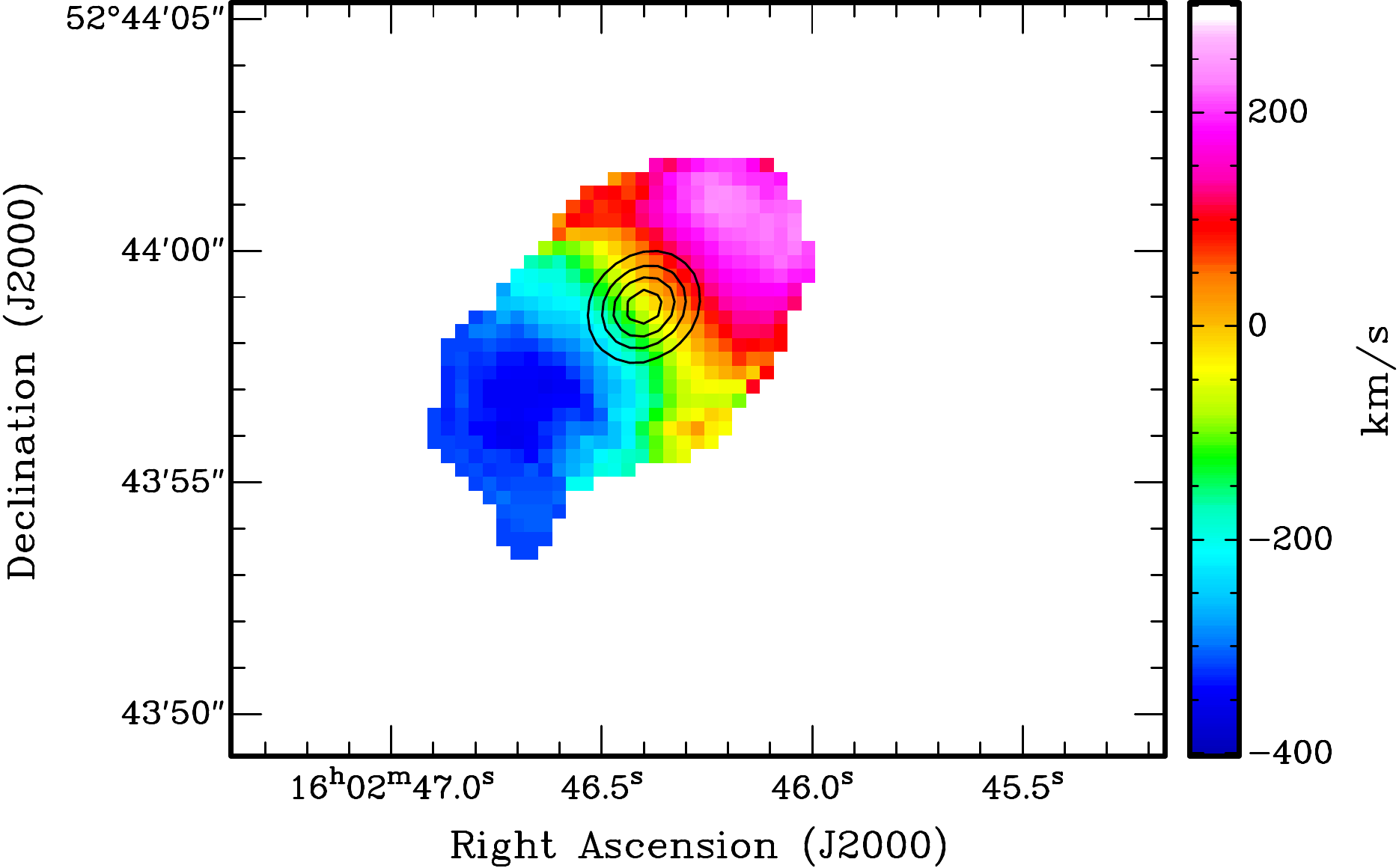} 
\includegraphics[height=0.3\textwidth]{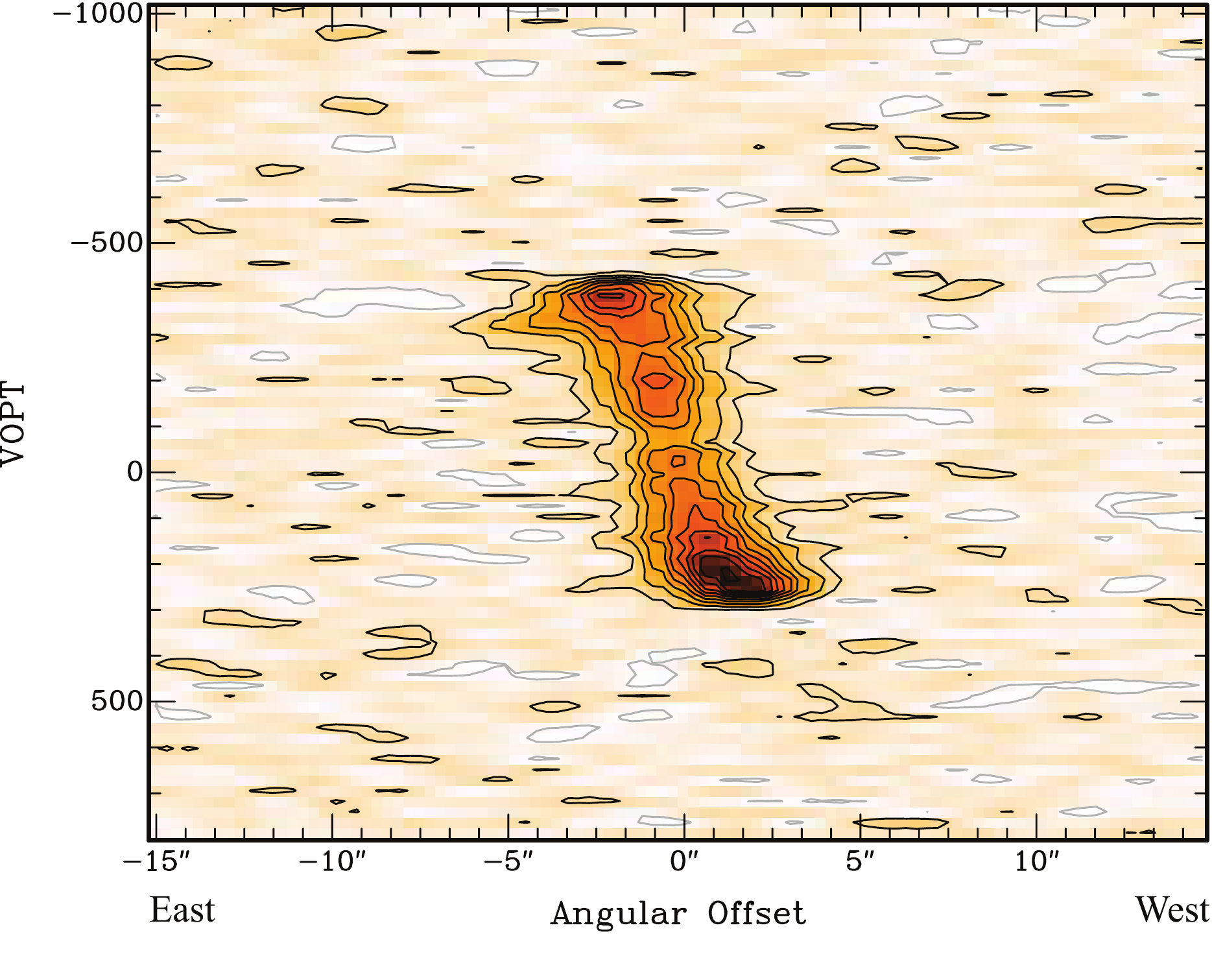}\end{adjustwidth}
\caption{\label{fig:4C52.37}Left/Centre: 4C~52.37 moment-0 and moment-1 maps obtained  using a mask to include only emission above 4-$\sigma$ level. This was made from the cube with robust weighting (see Table~\ref{tab:Observations}). Contours show the radio continuum map at a similar spatial resolution; contour levels are 2.5, 5.0, 7.5, and~10 \mJybeam. The inset shows the continuum at 1.4~GHz obtained  using VLBI (from~\cite{Schulz21}). Right: position--velocity plot along the major axis of the distribution of the molecular gas. The~small offset in the systemic velocity is likely due to the low accuracy of the redshift of this source. Contour levels are --0.5, 0.5, 1.5, ...,  8.5 \mJybeam.}
\end{figure}

The  optical  spectrum  of 4C~52.37 is characterised  by strong, broad permitted, and forbidden emission lines. This radio galaxy is known to be embedded in an \HI\-rich medium~\cite{Gereb15,Schulz21}, where the \HI\ appears kinematically disturbed. Broad, blueshifted \HI\ absorption was found  covering  more than 600 \kms\ \cite{Gereb15}. The~broad \HI\ absorption is mostly concentrated in the central regions (i.e.,\ the central $\sim$100 pc) of the radio source~\cite{Schulz21}. The~morphology of the radio continuum on VLBI scales is shown in the inset in \mbox{Figure~\ref{fig:4C52.37}}, whereas the mm continuum from the NOEMA observations is unresolved (contours in \mbox{Figure~\ref{fig:4C52.37}}).

For this object, our \coOne\ observations are of relatively low resolution with respect to the size of the radio source. This is because no previous observations of the molecular gas were available and, therefore, a conservative approach was used for our detection experiment. Higher resolution observations are needed for a more complete assessment of any jet--molecular gas~interaction. 


In the NOEMA observations, we detected CO(1-0) located in a regular disc of  $12\times 10^9$ \msun\ (assuming the Galactic value $\alpha_{\rm CO} = 3.6$ $M_\odot/({\rm K\ km\ s^{-1}\ pc^2})$) and distributed over a total extent of 10 arcsec ($\sim$19.4 kpc). The~gas reaches larger radii ($\sim$6 arcsec, 11.7 kpc) on the SE side. 
The moment maps and the position--velocity plot are shown in Figure~\ref{fig:4C52.37}.  In~particular, the position--velocity plot is characteristic of a regularly rotating disc. The~molecular disc is oriented in PA $129^\circ$, which is a 40$^\circ$ offset (in projection) from the radio axis (\cite{Schulz21}; see the inset in Figure~\ref{fig:4C52.37}). Considering the axis ratio of the integrated CO emission, the disc has a fairly high inclination (i.e\  about 60--70$^\circ$) but~it is not edge-on. 

As mentioned above, an~\HI\ outflow was detected in absorption against the central regions ($\sim$100 pc) of this source. Using a similar approach as for 3C~305, we can estimate a limit to the molecular outflow that could have been detected in our observations. We obtain about $8 \times 10^7$ \msun for~a 3-$\sigma$ profile covering an approximately 400 \kms\ width. If~the molecular outflow were located where the \HI\ outflow is found (between 60 and 150 pc in the E radio lobe), the~mass outflow rate of the molecular outflow would be less than $\sim$30 \msunyr. Compared to what was detected in \HI\ ($\sim$4 \msunyr;~\cite{Schulz21}), this suggests that  a massive molecular outflow is not present but~our observations are not sensitive enough to trace a more modest~outflow.

\subsection{4C~31.04}
\label{sec:4c31.04}

\begin{figure}[H]
\centering
\begin{adjustwidth}{-\extralength}{0cm}
\includegraphics[height=0.36\textwidth]{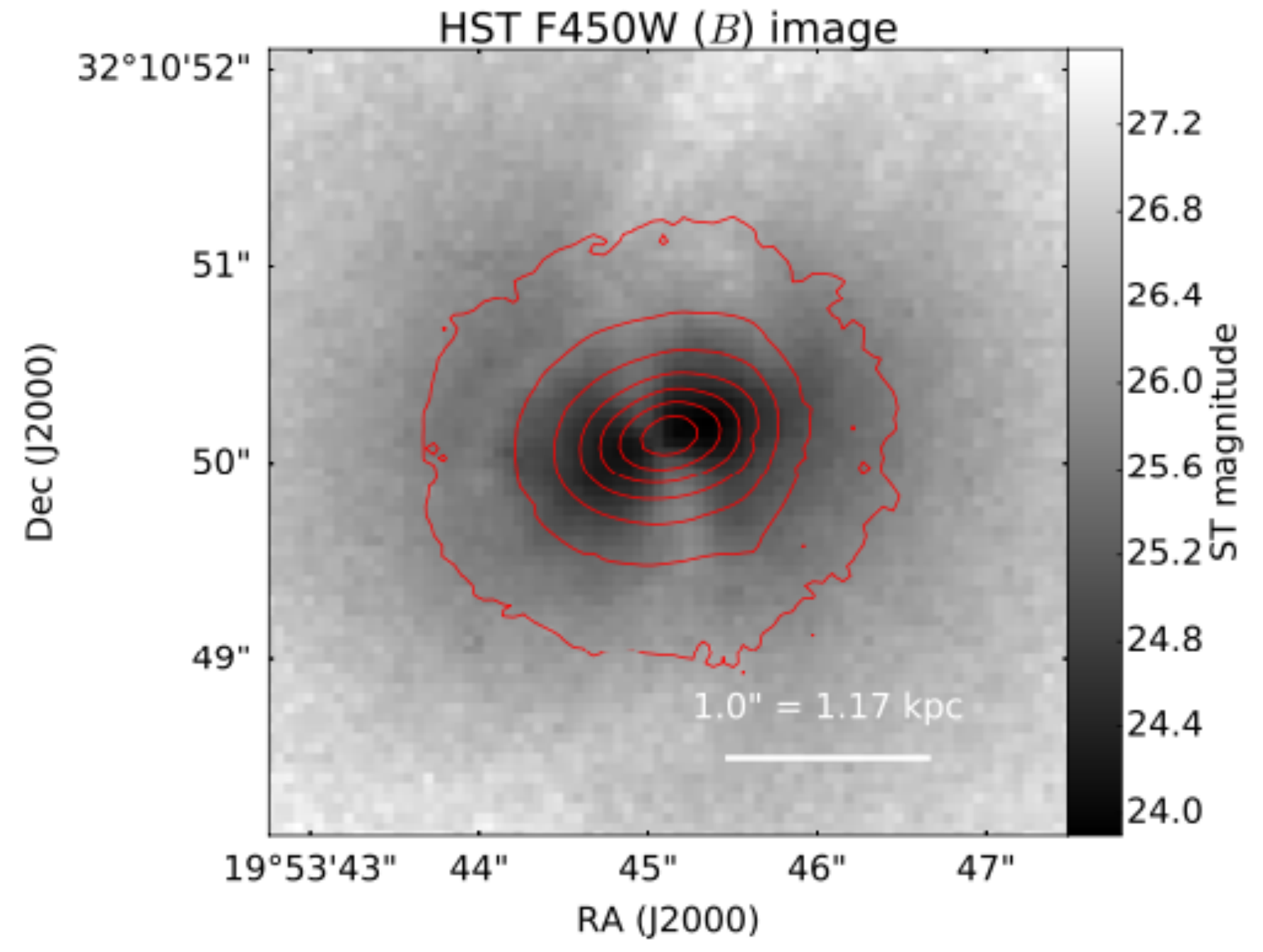}
\includegraphics[height=0.35\textwidth]{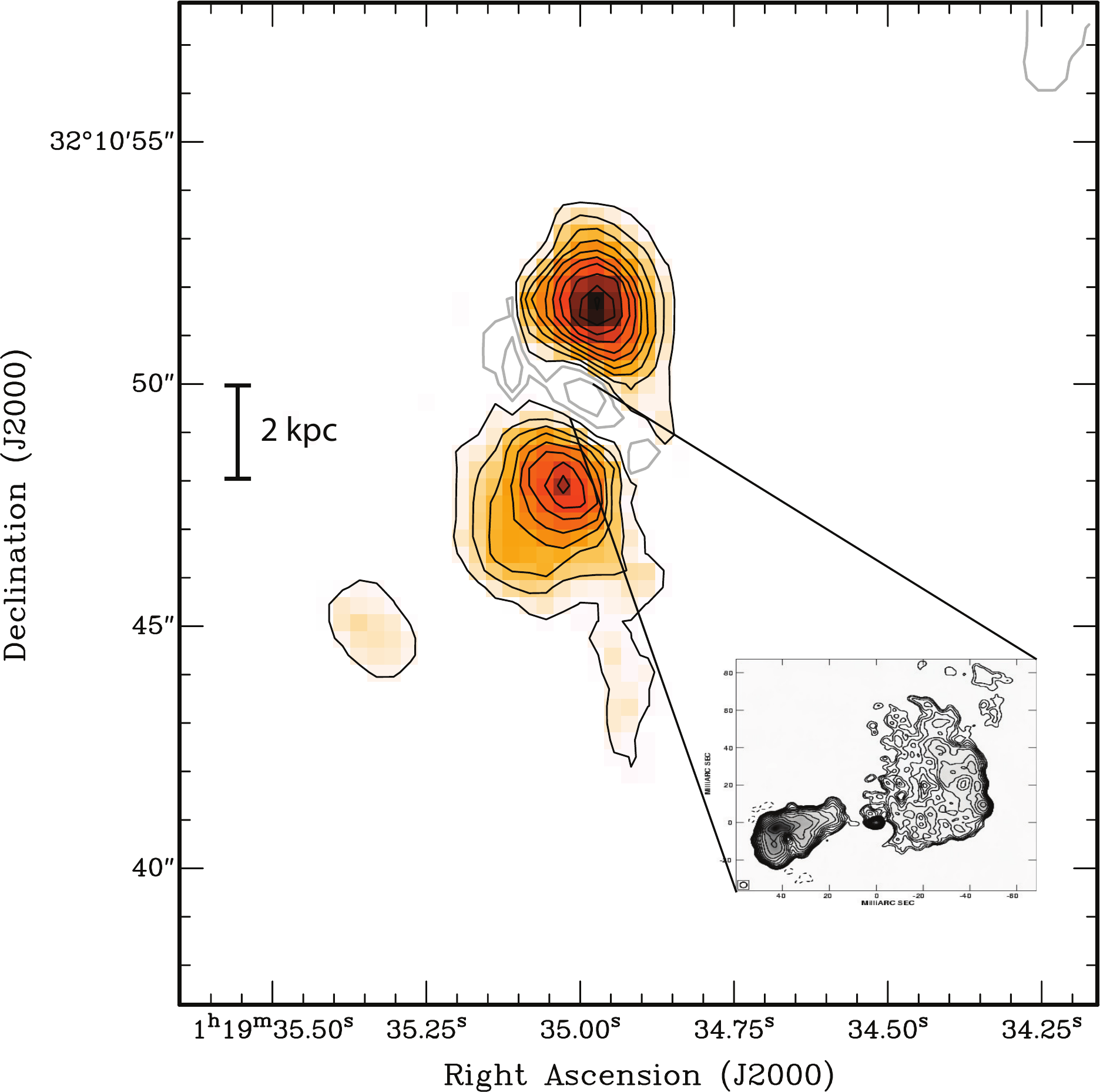}
\includegraphics[height=0.36\textwidth]{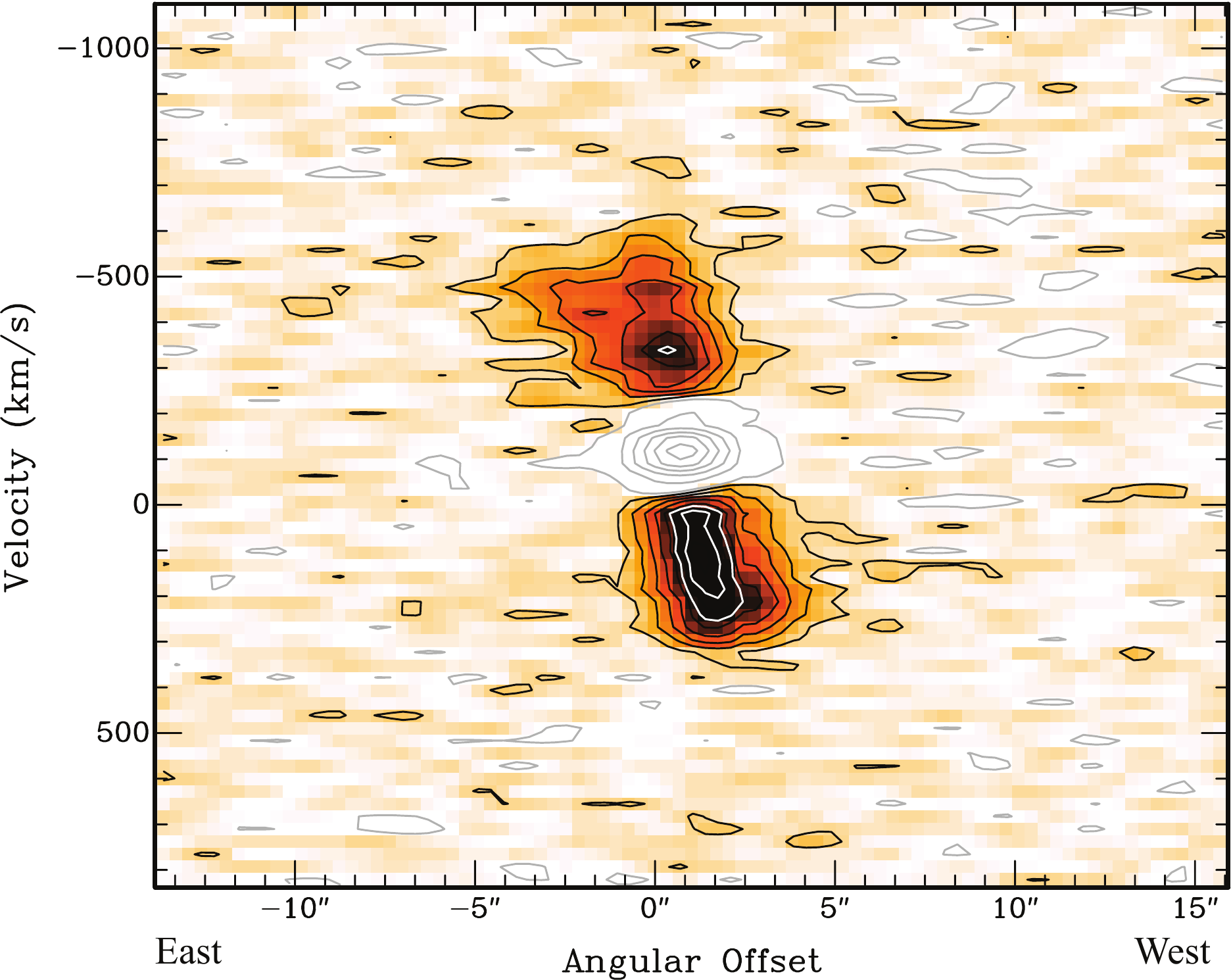}\end{adjustwidth}
\caption{\label{fig:4C31.04} Left: HST image of the dust lane in 4C~31.04 taken from~\cite{Zovaro19}; centre: moment-0 of the \coOne\ of 4C~31.04. The~position angle of the distribution of \coOne\ is consistent with that of the outer part of the dust lane. The~continuum image obtained with the VLBA at 5~GHz~\cite{Giroletti03} is shown in the inset. In~the central region, \coOne\ shows  strong absorption, which is indicated by the grey contours. Contour levels are --0.1, -0.01, 0.07, 1.07, ..., 8.07 Jy beam$^{-1}$\kms. Right: position--velocity plot along the major axis,  with~the absorption  indicated by light-grey contours. Contour levels are --26, --21, ..., --1, 0.5, 1.5, ... , 6.5   \mJybeam. The~large width of the absorption (full width at zero intensity $\sim$250 \kms) can be clearly seen. }

\end{figure}

This is a well-studied radio galaxy with a wealth of ancillary data~\cite{Perlman01,Struve12,Zovaro19}. The VLBI radio continuum emission shows a core and two lobes separated by only $\sim$100 pc (inset in Figure~\ref{fig:4C31.04} middle).   The lobes are very asymmetric, with the western lobe showing diffuse emission unlike the eastern lobe, which has a bright hotspot. The~host galaxy was observed with HST by~\cite{Perlman01} and shows an edge-on circum-nuclear dust lane with a prominent warp (Figure~\ref{fig:4C31.04} left). We observed 4C~31.04 both in \HI\ using VLBI and in \coOne\ with NOEMA. The~full presentation and discussion of the data can be found in the paper by Murthy~et~al.\ (currently being prepared), whereas here, we provide the highlights of the CO~observations. 


Molecular gas single-dish observations were carried out in \coOne\ and \coTwo\ by~\cite{Ocana10}, who detected $6.16 \times 10^8$ \msun\ of H$_2$ in emission and a strong absorption line in \coOne. Using IRAM Plateau de Bure (PdB),  \cite{Burillo07}  detected a circumnuclear disc of about 1.4 kpc in size in HCO+, both in emission and absorption. Although the kinematics of the gas are dominated by rotation, these data show evidence  of  distortions  and non-circular  motions,  which suggest  the disk is not  in a dynamically relaxed state and that it could provide the reservoir for fuelling the radio source. The~presence of a jet--ISM interaction is further supported by the observations of warm molecular gas~\cite{Zovaro19}. The~detection of [FeII] lines in the inner region has been described by these authors as tracing shocked gas ejected from the disc plane by a jet-blown bubble (300--400 pc in diameter), whereas the warm H$_2$ emission traces shock-excited the molecular gas in the  interior $\sim$1 kpc of  the  circumnuclear  disc~\cite{Zovaro19}.

Our NOEMA \coOne\ observations revealed a disc structure with a full extent of about 9 arcsec (10.4 kpc) and a mass of $2.4\times 10^9$ \msun\ (assuming the Galactic value $\alpha_{\rm CO} = 3.6$ $M_\odot/({\rm K\ km\ s^{-1}\ pc^2})$). This structure is shown in the middle in Figure~\ref{fig:4C31.04}. This disc follows the outer region of the dust lane seen by HST, as shown by the similar position angle. This complements the 1.4 kpc-sized HCO+ disc presented by~\cite{Burillo07}, which instead traces the inner regions. The~difference in position angle between this inner disc and the one observed in \coOne confirms that the warped structure observed in the dust lane is also traced by the cold molecular~gas. 

Strong absorption was seen against the 3 mm continuum, which is unresolved at our resolution. The absorption has a full width at zero intensity (FWZI) of $\sim$$250$ \kms and~is blueshifted with respect to the systemic velocity, as~can be seen on the right in Figure~\ref{fig:4C31.04}. Interestingly, similar properties were also found for the \HI\ absorption observed with  VLBI (Murthy~et~al., in~prep), which shows an FWZI of $\sim$ 300 \kms. However, although the CO continuum emission is unresolved (the size of the radio source is much smaller than the resolution of the present \coOne\ observations),  the~VLBI resolved the continuum emission and allowed us to locate the absorption and derive the resolved kinematics of the \HI\ gas. The \HI\ absorption is clearly extended, covering most of the radio lobes, and~blueshifted compared to the systemic velocity of the host galaxy. This indicates that it must be the result of kinematically disturbed gas not following the regular rotation of the large-scale disc. As~mentioned above, the study of the warm molecular gas also suggests the presence of a strong interaction and, in~particular, an expanding gas shell driven by the radio jets interacting with the ambient ISM. Thus, both the \HI\ and the cold molecular gas appear to be part of this structure. In addition, for this target as~for 4C 52.37, higher-spatial-resolution CO observations are needed to follow the molecular gas around the radio~source in more detail.

\subsection{CN(1-0) Lines}
\label{sec:CN}

\begin{figure}[H]
\includegraphics[width=0.5\textwidth]{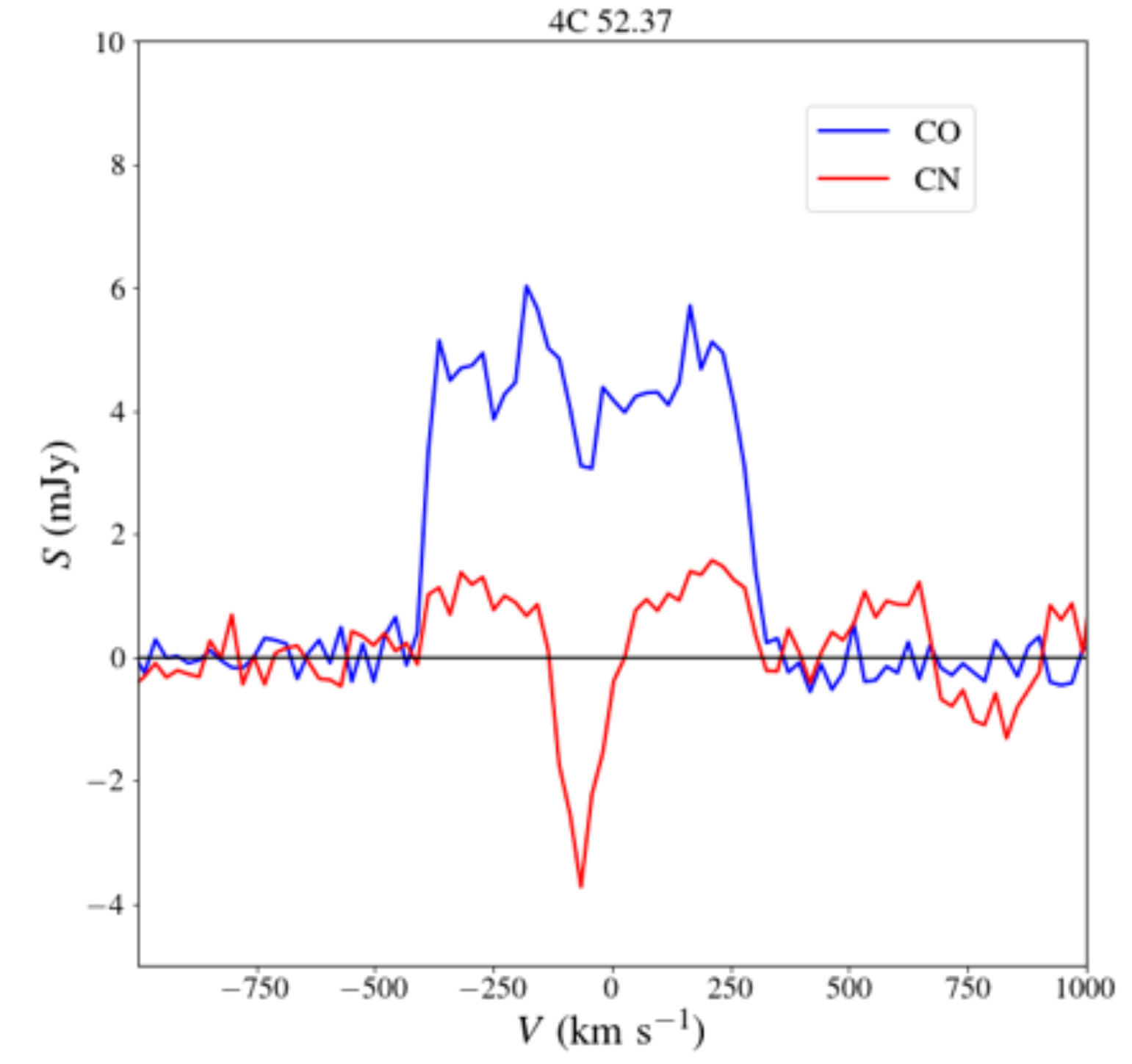}
\includegraphics[width=0.5\textwidth]{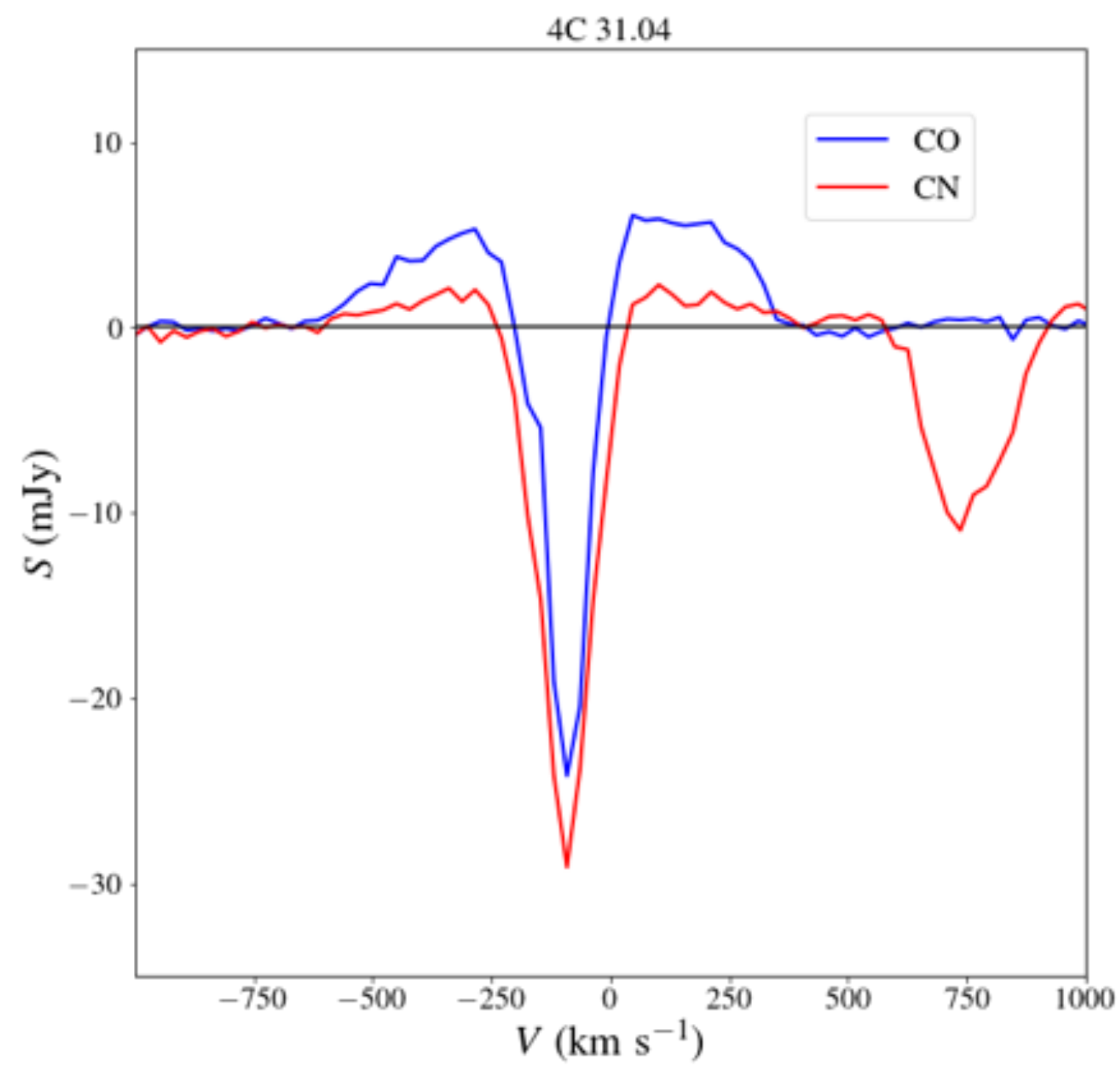}
\caption{\label{fig:CN} Integrated profile of \coOne\ and CN(1-0) for 4C~52.37 (left) and 4C~31.04 (right). In~both sources, the~two lines show both absorption and emission.
}
\end{figure}

The broad band of NOEMA allowed us to investigate the presence of the cyanide radical (CN) lines simultaneously with the \coOne\ lines. We detected the CN lines in two of the three objects, 4C 52.37 and 4C 31.04, and~we show the integrated spectra in Figure~\ref{fig:CN}. 

In the two sources, the~CN lines exhibit a mix of absorption and emission. In~4C~31.04, the~CN feature has the same structure as the CO, both in emission and absorption. However, the~CN(0-1) absorption is deeper than that of the CO because the CN molecule has a much larger dipole moment than CO, making the critical density of the corresponding CN(1-0) line larger by a factor of 10-100 than that of the CO(1-0) line. In~both sources, two sets of lines are seen in CN. In~4C~31.04, the~two absorption lines  have a peak line-strength ratio of about 2:1. The~deepest absorption line is due to the CN ($N$ = 0-1, $J$ = 3/2-1/2) group of hyperfine structure lines (unresolved here at a resolution of 24 km/s, centred at a rest frequency of 113.49 GHz), and~the weaker feature at a velocity of +750 km/s is the CN ($N$ = 0-1, $J$ = 1/2-1/2) group (centred around 113.16 GHz). This feature was also detected, although~weaker, in~4C~52.37.

CN molecules are primarily produced by photo-dissociation reactions of HCN in the moderately dense ($\approx$$10^3$~\cc) molecular phase of the ISM~\cite{Boger05}. Thus, CN emission lines mostly trace molecular gas irradiated by strong UV radiation fields. We estimated the integrated CO/CN ratios for the emission and absorption features using the brighter CN line, with~the caveat that the absorption could  be partly filled by emission. In~4C~52.37, we found a ratio of CO(1-0)/CN = 4 for the total integrated profile. In~4C~31.04, CO(1-0)/CN = 3 for emission and~CO(1-0)/CN = 0.8 for absorption.  A~quantitative exploitation of those line ratios can be found in an as-yet unpublished paper but here, we qualitatively discuss why those ratios are significantly lower than those in nearby galaxies, which have, on average, CO/CN $\approx 10$, with internal variations of a factor of 3-10 depending on the environment~\cite{Wilson18, Rose19}. 

Models suggest that the CN/CO abundance and line flux ratios are enhanced in low-density, irradiated molecular shocks (such as the CH+/CO ratio,~\cite{Godard19,Lehmann22}); high-velocity dispersion diffuse molecular gas~\cite{Wakelam15}; or X-ray-dominated regions~\cite{Meijerink07}. For~instance, high column densities of CN have  been found close to AGN~\cite{Riechers07, Rose19} and~elevated CN/CO line flux ratios have been  detected in the outflow of Mrk~231~\cite{Cicone20} or from an absorbing merging system tidal tail towards the background quasar G0248~\cite{Combes19}. Given the conditions originating from the kinematically disturbed gas in our sources, it is likely that one or a combination of some of these processes could be responsible for the elevated CN flux relative to \coOne\ in 4C~52.37 and 4C~31.04. A~more detailed morpho-kinematical study of line ratios compared with chemical and dynamical modelling could identify the dominant~mechanisms.

In 3C~305, we did not detect the CN lines. Considering the rms noise in the cube of $\sim$0.5 \mJybeam\ and the CO signal between 5 mJy (NE) and 4 mJy (SW), this gives a 3-$\sigma$ lower limit for the ratio of CO/CN of $\sim$3. Thus, this does not allow us to draw strong conclusions about the conditions in this~object. 

\section{Results from the Other Young Radio Galaxies in the~Sample}
\label{sec:publishedResults}

Here, we briefly summarise the main results obtained from the CO observations of the other objects in the sample, before~combining the results of all sources in Section\ \ref{sec:generalResults} and outlining the picture obtained so far. For a detailed discussion of the results, we refer the reader to the published papers listed in Table~\ref{tab:PropertiesTargets}.

\begin{figure}[H]
    \includegraphics[angle=0, width=0.7\textwidth]{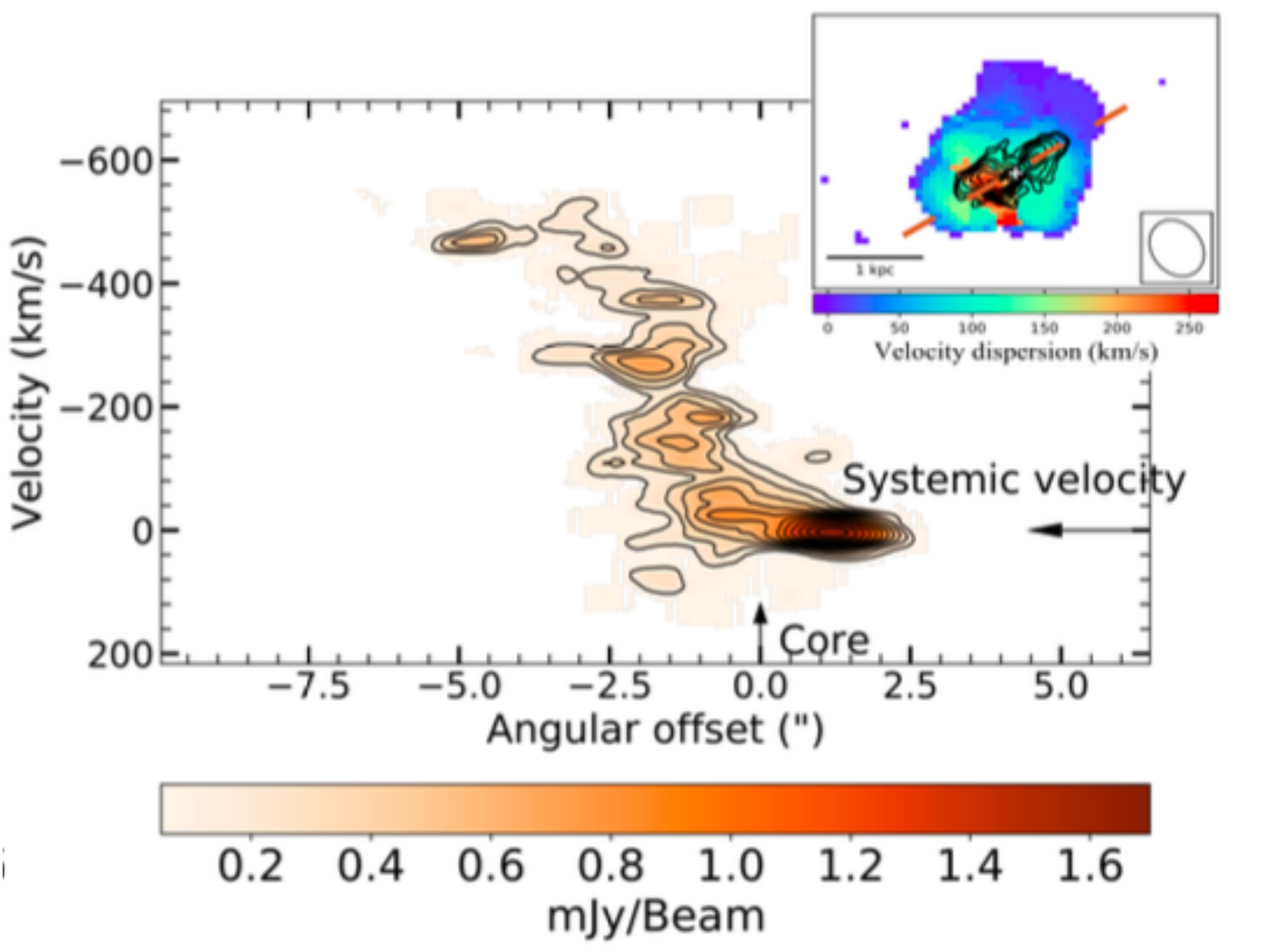}
    \caption{\label{fig:0258outflow}Position--velocity diagram of the circumnuclear gas in B2 0258+35 extracted along the radio axis, as shown in the inset. The~figure shows the velocity field of CO(1-0) superimposed with the contours of the radio continuum.  The~kinematics of the gas distinctly deviate from regular rotation and the outflow is offset to the southeast of the radio core. Figure taken from~\cite{Murthy22}.}
\end{figure}

The first object we discuss is B2 0258+35 ($L_{\rm 1.4~GHz} = 2.1 \times 10^{23}$ \WHz), highlighting the impact of relatively low-power radio jets, as~ presented in~\cite{Murthy22}. NOEMA CO(1-0) observations have shown that the circumnuclear molecular gas is highly turbulent and a fast (FWHM$\sim$350 \kms) jet-driven outflow of $\sim 2.6 \times 10^6$ \msun\ has been detected (Figure~\ref{fig:0258outflow} reproduced from~\cite{Murthy22}).  This outflow dominates the kinematics of the molecular gas in the central kpc and makes up $\sim$75\% of the total gas in the nuclear region. At~this rate, the~jet will deplete the kpc-scale molecular gas reservoir on a relatively short time scale of $\sim$$2 \times 10^6$~yr. This finding  confirms the predictions of the numerical simulations described in Section 1 and shown in Figure~\ref{fig:simulations}.

The results on B2~0258+35 confirm that low-power jets (in the case of this source, of about   $10^{44}$ \ergs~\cite{Murthy22}) have enough energy to drive outflows. Interestingly, this was also seen in the Chandra X-ray  observations of this source, which  revealed extended X-ray emission  both along and orthogonal to the jet~\cite{Fabbiano22}. At~the end of the main kpc jet, lower-energy X-ray emission is seen coinciding with the region where the molecular gas shows high turbulence and fast outflow motions. This suggests that the hot ISM may be compressed by the jet and that the molecular outflow can result from more efficient cooling, which is in agreement with the predictions of the numerical simulation of the jet--ISM interaction for low-power~jets.

It is interesting to note that for one of the best examples of a low-power radio jet impacting the molecular gas, IC~5063~\cite{Morganti15}, 
the Chandra X-ray  observations confirm the presence of strong shocks through the detection of localised Fe XXV emission in the region where the jet--ISM interaction occurs (see~\cite{Travascio21} for details). Thus, these very different phases of the gas (ranging from cold molecular to hot gas) appear to provide a consistent picture of the interplay between jets and the~ISM.

We also expanded our study of the molecular gas to young radio galaxies with high-power jets; the most extreme cases are PKS~1549--79~\cite{Oosterloo19} and PKS~0023--26~\cite{Morganti21-0023}. In~addition to powerful radio sources, these two objects also host a quasar-like optical AGN. Emitting  both radiative and mechanical energy, the~impact of the AGN is expected to be quite prominent in these~sources. 

In PKS~1549--79, using ALMA, we found the~most massive outflow of molecular gas in the sample ($\sim$650 \msunyr;~\cite{Oosterloo19}). However, the~outflow is limited to the inner 200 pc of the galaxy despite the presence of a powerful jet (about 300 pc in size) in combination with a powerful quasar AGN, with~bolometric luminosity between $6$ and $9.6 \times 10^{45}$ \ergs\ \cite{Santoro20}. A~circumnuclear disc of $M_{\rm H_2} =  2.6 \times 10^8$ \msun\ was also observed and appears to co-exist with the outflow. This would suggest a depletion time of $\sim$$10^5$ yr, although~a tidal tail of molecular gas, which was still in the process of falling in, was observed and may help to refuel the disc (and likely the AGN; see~\cite{Oosterloo19} for details).

PKS~0023--26 is also radio powerful but more extended, i.e.,\ it is one of the largest radio galaxies in the sample~\cite{Morganti21-0023}. The~radio emission  of this radio galaxy is  a few  kiloparsec in size, thus  probing  interesting  intermediate  scales where the young jet may start to break free from the inner dense regions of the ISM. The presence of direct interactions between the jet and the ISM was seen in the broad profiles of the CO observed in the inner kpc region. On~larger scales, the~molecular gas instead  appears to avoid the radio lobes and wrap around them. Although~the kinematics of the gas are disturbed, the~profiles are less broad than in the centre and the gas appears to be affected not by the direct interaction with the jet but~by the expanding cocoon surrounding the radio source, likely dispersing and heating pre-existing molecular clouds (see~\cite{Morganti21-0023} for details).

These results suggest an evolutionary sequence, similar to that predicted by the numerical simulations (e.g.,\ \cite{Sutherland07}) in~which the impact of the jet on the surrounding medium changes with its expansion. We revisit this in Section\ \ref{sec:discussionFuture}.

\section{The Picture So~Far}
\label{sec:generalResults}

From studying  the cold molecular gas around eight young radio galaxies, we found a variety of physical and kinematic properties of the observed CO structures, suggesting a complex interplay between jets and cold molecular gas. This was not unexpected given that it was shown by numerical simulations (e.g.,\  \cite{Mukherjee18}) and several parameters play a role in defining the impact of such an interaction. We also found some interesting trends, which can be compared with the predictions of the~simulations.

In most objects studied (with the possible exception of PKS~0023--26 and the central region of B2~0258+35), we saw a large fraction of the molecular gas distributed in rotating structures. The~full sizes of these structures cover a broad range from~sub-kpc to about 20~kpc. Similar sizes of molecular discs were also seen in evolved, large-scale radio galaxies and radio-quiet early-type galaxies (e.g.,\ \cite{Alatalo13,Boizelle17,Ruffa19,Ramos22} and Tadhunter~et~al., currently being prepared). Assuming it would take at least a few rotations to form such structures, the~discs we observed must have formed between a few $\times 10^6$ yr ago for~the discs of hundreds of pc and~ $\sim$$10^8$~yr ago for the largest (a few kpc) discs. They were, therefore, in~place before the radio source \mbox{(re-)started}. This suggests that, although~these structures can be impacted in various ways by a young jet (see below), they are seldom completely disrupted by the onset and evolution of the~jet, as illustrated by the simulations in Figure~\ref{fig:simulations}.

Our observations show, as seen previously for other phases of the gas, that young ($\lta$$ 10^6$ yr) radio jets are able to drive fast and massive gas outflows (see IC~5063,  B2~0258+35 and PKS~1549--79;~\cite{Morganti15,Murthy22,Oosterloo19} for the clearest cases). However, the~results from the cold molecular gas show, more clearly than in the case of the ionised gas, that low-power jets (i.e.,\ $P_{\rm jet}<10^{45}$ \ergs) are able to drive such massive outflows. This is the case for IC~5063 and B2~0258+35. This confirms the predictions of the simulations, which show that, although~ low-power jets can temporarily remain trapped in the ISM, they can accumulate enough energy during this time to break through the gas and drive outflows~\cite{Mukherjee18}. Because~they are less powerful, the~entire process takes longer than for powerful jets but can be extremely effective. This is relevant because~it highlights the relevance of the feedback of low-power sources, which are more common than powerful sources.   

However, all the observed molecular outflows are limited to the central (at most kpc) regions. This is also the case for objects (such as the two most powerful sources in the sample, PKS~1549-79 and PKS~0023-26) where both a strong radio and an optical AGN are present and, therefore, the~AGN impact is expected to be high. 
 
In two cases, the~depletion time of the outflows could also be derived.  For~B2~0258+35,  the~derived  mass  outflow  rate (between 5  and  10 \msunyr) implies that the kiloparsec-scale  molecular  gas  reservoir will be depleted in $\sim$$2 \times 10^6$ yr (see~\cite{Murthy22} for details).  
Even shorter is the depletion time found for PKS1~549--79, where the massive outflow would deplete the central disc on a time scale of only $\sim$$10^5$ yr. 
Although we are limited by the small number of objects studied, these results suggest that the conditions for developing an outflow could last only  a relatively short time, perhaps explaining the fact that molecular gas outflows are not observed ubiquitously.  The~outflow phase can, however, be recurrent in the cycles  of activity that are known to occur during the life cycles of radio galaxies (see~\cite{Jurlin20} and the references therein). 

In addition to outflows, we found  disturbed and highly turbulent gas in the regions co-spatial with the radio emission in most of the objects. This is another sign of the effect of the expansion of the young radio lobes into the surrounding~medium.  

A further indication of the impact of the expanding jet can be found in  the physical conditions of the gas. These appear to be different in the  kinematically disturbed gas from those in regularly rotating gas. We observed this, for~example, in~the CO line ratios in IC~5063 and PKS~1549--79 (the two objects where multiple transitions are available; see~\cite{Dasyra16,Oosterloo17,Oosterloo19}). The  line ratios from different  CO transitions were found to be substantially different in the kinematically disturbed gas from what is typically found in relaxed galaxy discs, with~the higher transitions being relatively much brighter in the disturbed gas. The~observed line ratios imply that the molecular gas near the AGN is optically thin and has elevated excitation temperatures ($T_{\rm ex} > 50$ K) as~a result of the impact of strong shocks. 
Similar results have been found by~\cite{Ruffa22} in the case of the young radio source hosted by NGC~3100. Using multiple CO transitions, these authors found a high-excitation molecular gas component in the region close to the radio jet, whereas the gas becomes progressively less excited at increasing distances from the radio structure. The~fact that in NGC~3100 only mild signatures of the interaction can be seen in the kinematics~\cite{Ruffa22} suggests that the line ratios can be a more sensitive indicator of the presence of jet--ISM interactions. 

Thus, to~obtain a complete picture of the impact of the radio plasma on the ISM, not only are the kinematics of the gas needed but~also  observations of multiple transitions (and/or multiple molecules) in order to derive the conditions of the gas such as  the~excitation temperature. This is also shown by the detection of CN lines in two of our targets. The~broad band of NOEMA allowed us to not only detect these lines but~also measure the CN/CO fraction. The high fraction detected compared to that typically found in quiescent gas discs can be explained as the result of the impact of the AGN on the conditions of the gas. More detailed investigations are required in order to fully quantify the~implications. 

One of the goals of this project was to investigate whether the interaction between the radio jet and the ISM changes as the jet expands.  Our results suggest that  this change indeed occurs. For~the largest objects in our sample (PKS~0023 --26 and 3C~305, both a few kpc in size), the~molecular gas appears to avoid the brighter part of the lobes and instead wraps around them. This is particularly clear in PKS~0023--26 in the channel maps shown in Figure\ 5 of~\cite{Morganti21-0023} and a similar trend was seen in 3C~305. 

Because of this change (albeit based on only a few cases), we propose a possible evolutionary scenario (see also Figure\ 11 in~\cite{Morganti21-0023}). The fast and massive outflows are limited to the very inner regions close to the centre, where the newly emerged jet interacts directly with the gas clouds while opening its way in the surrounding ISM. When the jets reach the kpc radii, this is replaced by a milder, more gentle interaction driven by the expanding (both along and orthogonal to the jet) cocoon created as a consequence of the initial direct jet--cloud interaction. This results in the dispersing and heating of the ISM. These findings, which are in line with  predictions from numerical simulations of jets interacting with a clumpy medium, also suggest that the effect of expanding shells of shocked gas---reminiscent of the ``maintaining mode'' of feedback associated with X-ray cavities in cluster galaxies---could provide an important way for radio plasma jets to affect the galactic~ISM.


\begin{figure}[h]
\includegraphics[width=0.65\textwidth]{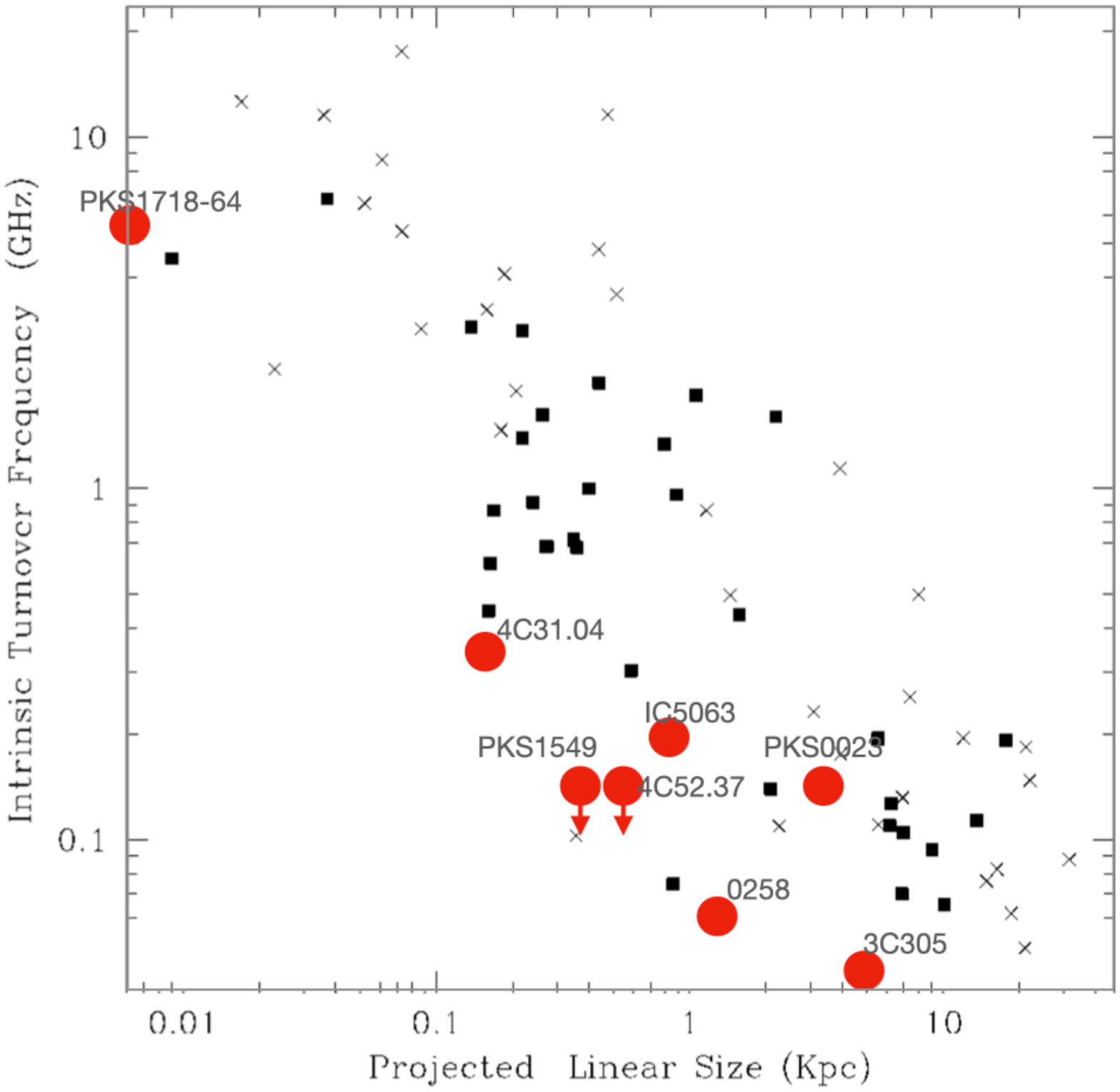}
\caption{\label{fig:turnover} Turnover frequencies of our targets (red circles) compared to CSS/GPS from the literature (black points). Original plot taken from~\cite{ODea98}.}
\end{figure}

In our study, we have seen several signatures of interaction between the radio plasma jets and the ISM in the vast majority of young radio sources located in a gas-rich environment. This kind of interaction may, in~turn, also affect the general evolution of the radio jets.  A~relationship between the turnover frequency of the radio spectrum and the size of the radio source has already been suggested in early studies of young radio galaxies (i.e.,\  \cite{Fanti90,ODea98}). This relationship has been explained by the turnover frequency being due to synchrotron self-absorption or free--free absorption due to a dense medium and it evolves as the source grows. It is interesting to see that the turnover frequencies of the sources studied here (in some cases, derived by adding new low-frequency points, e.g.,\ from GLEAM or LOFAR, see, e.g.,\ \cite{Brienza18,Morganti21-0023}) are all located below the correlation-size turnover frequency from the literature, as~shown in Figure~\ref{fig:turnover}. The most likely explanation for this is that  radio  sources that strongly interact with a rich medium are affected in the speed of their expansion, resulting in a smaller size given the turnover frequency observed. This would suggest that the expansion of the radio lobe has been slowed down due to the interaction with the rich~medium.  

As a final remark, it is interesting to note that the only object where we observed clear signs of molecular gas in the process of falling into the SMBH was the smaller radio source in the sample, PKS~1718-63. With~only one object available, we can only speculate that the very young age of the jet means that the jet--ISM interaction is yet to develop. Thus, in~this object, we can more clearly see the gas that is instead in the process of feeding the AGN before the disentangling of the two processes (feeding and feedback) becomes too~complex.

\section{Discussion and Future~Work }
\label{sec:discussionFuture}

We explored the jet--ISM interplay for a small sample of young radio galaxies in which the newly born  jet is expanding into the rich ISM of the central regions of the host galaxy. We found that in the vast majority of the objects (with perhaps one exception), the~gas is kinematically disturbed in the region co-spatial with the radio galaxy, showing the coupling between the jets and the~ISM.

In only a small fraction of the sources were the observed kinematics of the molecular gas extreme.  However, the~finding that radio jets and, in particular, low-power jets, are able to produce fast and massive outflows, makes this type of AGN relevant for~cosmological simulations, even on galactic scales. Indeed, some recent cosmological simulations have started including more realistic descriptions of the impact of the jets~\cite{Talbot22} despite the huge challenges involved. We further observed that gas outflows are limited to the very inner regions of the galaxy. Thus, the~outflows observed (even in powerful AGN) do not seem to be large or extended enough to provide feedback and affect star formation on galactic~scales. 

Interestingly, in~the largest radio sources (a few kpc in size), we observed  the gas avoiding the region with the radio continuum emission and wrapping around the radio lobes. Thus, we suggest an evolution in the type of interaction between radio plasma and gas. Outflows from the direct jet/cloud interaction in the clumpy medium are occurring in the inner region (few hundred parsecs/sub-kpc), whereas on larger scales (larger than a kpc), the impact is mediated by the cocoon of the shocked gas created by the jet--cloud interaction; this cocoon pushes  the gas aside and heats it. Consistent with this is that a number of our targets observed with Chandra show signatures of  X-ray gas thermally heated by the interaction between the radio plasma and the~ISM. 

If this scenario is correct, we argue that the impact of the radio plasma will continue on galactic scales as the jets expand. The~impact of galactic-scale jets (GSJ;~\cite{Webster21a,Webster21b}) has already been  investigated in a handful of objects, with~observations looking at the distribution of X-ray gas. Using the Chandra observations of these sources,~\cite{Croston07,Croston09,Mingo11,Mingo12} have found the presence of structures of hot gas thermally heated by shocks from the jets, as shown by the morphological correspondence and distribution around the radio lobes. 
Thus, in~this phase, the role of the radio plasma could mostly be preventing the cooling of the gas, similar to what occurs in  the hot medium in clusters by  the creation of X-ray cavities.  A~detailed study of the impact of GSJ   on both the hot and cold/warm gas could be the way forward  to obtain a full overview of the role of the radio plasma jets in~feedback. 

The trends we have found for a small number of objects need to be confirmed by observations of larger samples. Although~this is a demanding task requiring high spatial resolution and sensitive observations with oversubscribed  telescopes such as ALMA and NOEMA, the continuously expanding ALMA archive will provide a growing database, which will hopefully soon allow for the expansion of the present results.

\vspace{6pt} 



\authorcontributions{Conceptualisation and formal analysis: R.M., S.M., P.G., T.O., and S.G.-B.;  data curation, S.M. and T.O.;  writing---review and editing, R.M., S.M., P.G., T.O., and S.G.-B. All authors have read and agreed to the published version of the~manuscript.}

\funding{This research received no external~funding.}

\dataavailability{The data from the NOEMA archive and the final calibrated data, images, and cubes are available on request to the authors. }

\acknowledgments{
IRAM is supported by INSU/CNRS (France), MPG (Germany), and IGN (Spain). We thank the IRAM staff for making these observations possible. S.M. thanks Orsolya Feher for help with data reduction. P.G. thanks the Centre National d'Etudes Spatiales (CNES), Pierre and Marie Curie University, Institut Universitaire de France, "Programme National de Cosmologie and Galaxies" (PNCG), and "Physique Chimie du Milieu Interstellaire" (PCMI) programs of CNRS/INSU for their financial support. The~IRAM NOEMA Interferometer data were reduced with the GILDAS, accessed on 8, 9 and 10 March 2021
 (\url{https://www.iram.fr/IRAMFR/GILDAS}). }

\conflictsofinterest{The authors declare no conflicts of~interest.} 

\sampleavailability{Samples of the compounds are available from the authors.}


\abbreviations{Abbreviations}{
The following abbreviations are used in this manuscript:\\

\noindent 
\begin{tabular}{@{}ll}
NOEMA &  Northern Extended Millimeter Array \\
ALMA & Atacama Large Millimeter Array \\
VLBI & Very-Long Baseline Interferometer \\
\end{tabular}
}

\begin{adjustwidth}{-\extralength}{0cm}
\printendnotes[custom] 

\reftitle{References}

\PublishersNote{}
\end{adjustwidth}
\end{document}